**Connecting Harbours. A comparison of traffic networks across ancient and medieval Europe**

Johannes Preiser-Kapeller, Österreichische Akademie der Wissenschaften, Institut für Mittelalterforschung, Abteilung für Byzanzforschung
Email: Johannes.Preiser-Kapeller@oeaw.ac.at

Lukas Werther, Friedrich-Schiller-Universität Jena, Seminar für Ur- und Frühgeschichtliche Archäologie
Email: lukas.werther@uni-jena.de

**Summary**

Ancient and medieval harbours connected via navigable and terrestrial routes could be interpreted as elements of complex traffic networks. Based on evidence from three projects in Priority Programme 1630 (*Fossa Carolina, Inland harbours in Central Europe* and *Byzantine harbours on the Balkan coasts*) we present a pioneer study to apply concepts and tools of network theory on archaeological and on written evidence as well as to integrate this data into different network models. Our diachronic approach allows for an analysis of the temporal and spatial dynamics of webs of connectivity with a focus on the 1st millennium AD. The combination of case studies on various spatial scales as well as from regions of inland and maritime navigation (Central Europe respectively the Seas around the Balkans) allows for the identification of structural similarities respectively difference between pre-modern traffic systems across Europe. The contribution is a first step towards further adaptions of tools of network analysis as an instrument for the connection and comparison of data across the projects of Priority Programme 1630.

# #H1#Harbours and Network Models – an Introduction

Over the last years, we observe an increasing number of studies which aim to survey, map and analyse the connections and interactions between various elements (individuals, groups, settlements, polities, but also objects or semantic entities) documented in historical or archaeological evidence with the help of network models in the form of graphs with "nodes" and "ties". Also a workshop within the Priority-Programme 1630 "Harbours from the Roman period to the Middle Ages" was devoted to this topic in Mainz in 2013, with the proceedings published in 2015.[1] Equally at the plenary meeting in Jena in January 2015, network theory has been discussed as one possible methodological approach allowing for a connection and comparison between various regions and periods integrated into the Priority-Programme 1630, leading to the establishment of an initial "working group" on this question, of which the current paper is a first outcome.

In general, network theory assumes "not only that ties matter, but that they are organized in a significant way, (and) that this or that (node) has an interesting position in terms of its ties."[2] Data on the categories, intensity, frequency and dynamics of interactions and relations between entities of interest is systematically collected in a way which allows for further mathematical analysis. Once a quantifiable network model has been created, it allows for a structural analysis on the level of single nodes with regard to their relative "centrality", on the level of groups of nodes with the aim to identify clusters of more closely interconnected nodes and at the level of the entire network with general parameters such as size or density enabling comparisons between networks.[3]

# #H1#Case Studies River Networks

Within the projects "Studies of inland harbours in the Frankish-German Empire as hubs for European communication networks (500-1250)" and "Fossa Carolina" and in cooperation with the working group "Data integration", we have created a database of harbours and landing sites at rivers and lakes in Central Europe, France and Northern Italy from the Roman period to the year 1250, also integrating the recently published catalogue of Christina Wawrzinek.[4] In order to integrate harbours in written sources we use an open harbour concept, which does e. g. also include documented starting points and destinations of transports by ship. The database is work in progress and the data acquisition status is still heterogeneous and somehow patchy, but nevertheless it enables diachronic and supra-regional perspectives which go far beyond hitherto existing possibilities. In this study, we use selected data from the river catchments of Danube, Po, Rhine and Rhône for the modelling and analysis of traffic networks and a comparison of their structure in the Roman (1st-5th cent. CE) and post-Roman (6th-early 11th cent. CE) period[5]. (**Fig. 1**) The selected rivers and their tributaries represent major axis of communication between the Mediterranean and Northern Europe on both sides of the main European watershed. Furthermore, they form a crucial link between the Scandinavian and Byzantine research clusters within the priority programme.

---

[1] Preiser-Kapeller/Daim 2015.

[2] Lemercier 2012, 22. Cf. also Brughmans 2012, the contributions in Knappett 2013, Collar/Coward/Brughmans/Mills 2015, and Preiser-Kapeller 2015a/b, for an overview of concepts and tools as well as further bibliography.

[3] Collar/Coward/Brughmans/Mills 2015, 17-25, includes also a most useful glossary of basic terms and concepts of network analysis as well as a well-balanced discussion of potential and pitfalls of network models in archaeology. For historical studies the best discussion in this regard is Lemercier 2012.

[4] http://www.spp-haefen.de/en/projects/binnenhaefen-im-fraenkisch-deutschen-reich/ resp. http://www.spp-haefen.de/en/projects/fossa-carolina/ resp. http://www.spp-haefen.de/en/home/ and http://www.spp-haefen.de/de/das-schwerpunktprogramm-1630/datenzusammenfuehrung/. Cf. Wawrzinek 2014.

In our model, harbour sites documented in historical or archaeological evidence serve as nodes and routes between them via river or lake navigation as links. Such models have been among the first applications of network theory to phenomena of the past; already in the 1960s and 1970s, F. R. Pitts created models of the medieval river trade network of Russia.[6] But in contrast to these first attempts, links in our model are both weighted (meaning that a quantity is attributed to them) and directed (a link leads from harbour A to harbour B, for instance). The aim is to integrate at least some aspects of what Leif Isaksen has called "transport friction" into our calculations; otherwise, the geographical realities and actual costs of communication and exchange between sites, which influenced the frequency and strength of connections, would be ignored in network building. Links are weighted by using the inverted geographical distance between them; thus, a link would be the stronger the shorter the distance between two nodes ("distant decay" effect).[7] We used the European CCM River and Catchment Database to document those distances in ArcGIS[8]. Real historical distances may differ significantly in some cases due to changes of river courses, but there is no workable alternative to date on a supra-regional scale[9]. The flow direction of a river effectuates easier and therefore stronger connections downstream. Therefore, directed links leading upstream (from harbour A to harbour B) are weighted with a third of the strength of links leading downstream (from harbour

---

[5] We surveyed this catchments systematically based on publications. The complete dataset will be published at the end of the ongoing project.

[6] Pitts 1978, 285-292. For the analysis of transport and traffic networks cf. Rodrigue/Comtois/Slack 2013, 307-317; Taaffe/Gauthier, Jr. 1973, 100-158; Ducruet/Zaidi 2012, 151-168. For a more general approach see Barthélemy 2011, 1-101. For transport networks of the past see also Carter 1969, 39-56; Gorenflo/Bell 1991, 80-98; Preiser-Kapeller 2015b.

[7] Isaksen 2008; Preiser-Kapeller 2015b. We documented this relationship in simple EXCEL tables, which are the basis for all further analysis.

[8] http://ccm.jrc.ec.europa.eu/php/index.php?action=view&id=23.

[9] Cf. Wawrzinek 2009, 171–180.

B to harbour A), which seems to be a feasible mean value rooted in written sources.[10] To improve this very simplifying model for future research, friction factors like individual river-specific weights for upstream-downstream ratios, river inclination, flow rate or navigable width could be integrated for particular hydrological units.

 On this basis, we modelled two different networks, for the Roman (**Fig. 2**) and Post-Roman (**Fig. 3**) period. As there is no connection between the four river catchments of Po, Rhine, Rhône and Danube on the waterway, we created four separate Roman sub-networks and four separate post-roman sub-networks in the first step.

## #H2#The Harbour Network in the Po Catchment

Our case study in the Po catchment illustrates elementary facets of the network analysis. We modelled two networks, one on the basis of data for the 1st-5th cent. CE (period I) and one for the 6th-early 11th cent. CE (period II) and determined the standard centrality measures on the level of individual nodes (degree and especially betweenness and closeness) and of the entire network. (**Tab. 1**) In the following we focus particularly on two centrality measures, which need to be explained in their significance for harbour networks: Betweenness centrality is equal to the number of shortest paths from all harbours to all other harbours that pass through a certain harbour. A harbour with a high betweenness centrality has a large influence on the transfer of ships and goods. It has a high potential for intermediation and could be understood as a "bottle-neck". Therefore, if a harbour with a high betweenness centrality is abandoned, the network may easily break apart. Closeness centrality on the other hand is a measure of reach. It measures the length of all paths between a harbour and all other harbours. The "closer" a harbour is, the lower is its total and average distance to all other harbours. Closeness also shows us, how fast it would take to spread ships, goods or information from a certain harbour to all other harbours in the network.

While a visualisation of the nodes (sized according to their relative centrality) on a geographical map (**Fig. 4**) illustrates continuities and changes with regard to the focal points of connectivity in these two models, a comparison of quantitative measures indicates significant differences in range, connectivity and complexity between the two graphs: The number of nodes is 25 % smaller in the period II-network, the number of links is only half the one of the period I-model. Period II-network´s (weighted) density is only two thirds and its clustering coefficient half the size as for the period I-model. Significant is also the difference of values for transitivity (which indicates the percentage of link pairs in the network where when node A is linked to node B and node B is linked to node C also node A is linked to node C) between period I (0.657) and period II (0.25). Another measure especially developed for transport networks (as planar graphs) is circuitry, measuring the share of the maximum number of cycles or circuits (= a finite, closed path in which the initial node of the linkage sequence coincides with the terminal node) actually present in a traffic network model and thus indicating the existence of additional or alternative paths between harbours in the network and its relative connectivity and complexity[11]; here the difference is even more significant (0.38 for the period I-model and 0.10 for the period II-model). The model for the riverine network in period II is thus not only spatially more confined, but also less well connected and complex when compared with the model for period I. This would correlate with widespread assumptions on the relative reduction of organisational, economic and infrastructural complexity from the Roman to the post-Roman period.[12] Nevertheless, we must consider that the data quality is still very heterogeneous due to different research foci with a special emphasis on the roman period[13]. Therefore it is to be expected that especially the patchy network of period II will show profound changes as soon as research intensifies.

---

[10] Isaksen 2008; Graßhoff/Mittenhuber 2009. See also Leidwanger/Knappett et al. 2014. For transport friction cf. also van Lanen et al. 2015, 144–159. For navigation upstream and downstream cf. Campbell 2012, 200-219; Suttor 2006, 181; Escher/Hirschmann 2005, 78-92; de Soto 2013, 1551–1576; Eckoldt 1980a; Ellmer 1984, 253-254. See also http://orbis.stanford.edu/orbis2012/#rivertransport.

## #H2#The Harbour Networks of Rhône, Rhine and Danube

With the same method, we created similar network models of riverine transport for the river systems of the Rhône, the Rhine and the Danube (until the territory of modern-day Austria).

A visualisation of the nodes (sized according to their relative centrality) on a geographical map again highlights continuities and especially changes with regard to the focal points of connectivity in the models for the two periods in all river systems. **(Fig. 5, Fig. 6)** The models for the Rhône-river system (**Fig. 7, Fig. 8**) show a profound decline of connectivity and complexity from period I to period II very similar to the Po catchment. Especially striking is the discontinuation of the major roman transit route from the Upper Rhône to the Swiss Lakes in period II.[14] Nevertheless, selected harbour sites like Lyon show strong continuities and remain focal points of connectivity in the Middle Ages.[15] For the Rhine-river system, the model for period I (**Fig. 9**) shows a relatively equal distribution of betweenness centrality among nodes along the entire Rhine, while in the model for period II (**Fig. 10**), high betweenness centrality values are clustered in the region between Speyer and Koblenz, with Mainz having the highest centrality, indicating an increased significance of the Main and its confluences as route of transport in the model for period II. This is also reflected in a significantly higher betweenness centralisation in the period II-model than in the period I-model, which differs from the results of our comparison of the two models for the Po- and the Rhône-network. Equally in contrast to the models for the Po and also the Rhône-network and despite of a (relatively small) decline in the number of nodes and links, measures of network complexity (clustering coefficient and especially transitivity and circuitry) of the model for period II for the Rhine are higher than in the Period I-model[16].

[11] Rodrigue/Comtois/Slack 2013 310, 313 and 315-316; Taaffe/Gauthier, Jr. 1973, 104-105: the circuitry or alpha-index is calculated as share of the maximum number of circuits actually present in a graph.
[12] Wickham 2005, for Italy especially 728-741; Szabó 1984, 125. For aspects of environmental change in this region cf. Cremonini/Labate/Curina 2013, 162-178; Patitucci Uggeri 2005. On the Po catchment in period I in general Campbell 2012, 302-309.
[13] Cf. Sindbæk 2015, 112-113.
[14] Cf. Delbarre-Bärtschi/Hathaway 2013.
[15] Cf. Schmidts 2011; Rossiaud 2007; Rieth 2010; Campbell 2012, 270-271.

## Interconnected Harbour Networks and Newman Cluster

In order to capture the integration of the four separate river systems of period 1 into one interconnected traffic system through the road network in the Roman period, we created a further (partially hypothetical) model (**Fig. 11**), which not only added the roads between selected connection points to the respective river systems, but also a canal between the rivers Moselle and Saone, which was planned during the Roman period in order to connect the traffic networks of Rhône and Rhine.[17] Terrestrial links are weighted by using the approximated road distance between them. We used the digital Roman Road Network of McCormick et al. to document those distances in ArcGIS.[18] The emerging combined network model of riverine and land routes is of course characterised by a higher circuitry measure than the separate network models; central points of connectivity (with regard to betweenness) are the intermediary hubs for the connections between Rhine and Rhône respectively across the Alps, with a barycentre towards the South. The upsurge of centrality in the combined network model for the harbours of Besancon, Toul, Lausanne, Yverdon, Bregenz and Augsburg is remarkable. Exactly these harbours work as points of transhipment between water and land to cross the watersheds between two or more hydrological systems. Along similar lines, we also created alternative interconnected models for period I (**Fig. 12, Fig. 13**) and period II (**Fig. 14, Fig. 15**) limited to the river systems of Rhine and Danube. The period I network integrates selected roman road connections bridging the watershed between Augsburg/*Augusta Vindelicum* respectively Burghöfe/*Summuntorium* and Pforzheim/*Portus*, between Augsburg and Bregenz/*Brigantium* and between Augsburg and Bad Zurzach/*Tenedo*. The period II network simulates the effects of a connection bridging the main European watershed through a working canal in the Carolingian period (the fossa Carolina/Karlsgraben) or rather a road corridor at the same position. A view on the betweenness centrality values respectively a comparison with the values for the two separate network models show the emergence of a new central intermediary zone along the axes of connection between the two river system but also that such a modification even more augments the central position of the already focal zone around Mainz. Regardless of the question whether the Fossa Carolina was finished or terrestrial routes formed the backbone of transportation across the watershed, the link between Rhine and Danube via several harbours at the rivers Main, Regnitz, Schwarzach and Altmühl is a fundamental development of the period II network. Compared with the period I interconnected network of all four river catchments, the "Roman" intermediary zone between the catchments of Rhône and Rhine with harbours at the very end of the navigable headwaters of Doubs, Rhône, Rhine and Moselle close to the main European watershed seems to be replaced by a new "Early Medieval" navigation corridor between Rhine and Danube.

---

[16] For the Rhine corridor cf. Kennecke 2014; McCormick 2001, 653-655; Campbell 2012; Johanek 1987, 7–68. .
[17] Tac. ann. 13, 53. Cf. Eckoldt 1980b, 29-34; Campbell 2012, 223-224.
[18] McCormick et al. 2013.

In order to get an idea about the inner structuring of the interconnected networks in different transport zones, we used the Newman-algorithm for the localisation of clusters, meaning the existence of groups of nodes more densely connected to each other than to the rest of the network respectively groups of harbours which could ship to each other on comparable costs[19]. The Newman-algorithm for the interconnected network model of all four river catchments identifies 15 regional transport zones (**Fig. 16**).[20] An intermediary cluster (no. 7) crossing the watershed integrates nodes from both along the Moselle and the Saône.

Once more, the Newman-algorithm for the interconnected model of period II limited to the river systems of Rhine and Danube and connected by the Fossa Carolina (**Fig. 17**) identifies no distinct intermediary cluster integrating nodes from both sides of the watershed. The harbours of the "Danube cluster" (no. 2) and the "Main-Regnitz cluster" (no. 1) are not connected densely. Some of the traffic zones in (**Fig. 16**) and (**Fig. 17**) or especially borderlines between them show continuities between period I and II, for example the separation of the Rhine-Main-confluence transport zone and the Upper Rhine and Bodensee transport zone or the continuity of the densely connected Danube transport zone.

The identification of fluvial transport zones on a regional scale with the help of the Newman algorithm may be a fruitful starting point for future research on the role of harbours as nodal points for the distribution of goods and people as well as on the use of different types of ships in the Roman and Medieval period. As a case study, we compared the period II Newman cluster-induced transport zones with the distribution of the well-known "Mayener Keramik" of the

---

[19] Newman 2010, 372-382; Preiser-Kapeller 2013; Preiser-Kapeller 2015 a/b.
[20] For this aspect see especially Westerdahl 2000, 11–20.

Early and High Middle Ages (**Fig. 18**), which is significantely clustered along waterways and therefore thought to be transported mainly on ships.[21] It is striking, that the ceramic distribution from the Mayen workshops to the South does not cross the borderline between Newman cluster 4 and 3 south of Speyer – one may assume, that the effort of transport upstream reached a critical point here. The same could be observed in the East, where the distribution of "Mayener Keramik" reaches the border zone between Newman cluster 1 and 7 at Karlburg and Würzburg, but no harbours further upstream.

## Integrating ORBIS

A more appropriate modelling of (inland) transport networks would of course necessitate the consideration of land routes and passages from riverine to maritime transport (as at the mouth of the rivers Rhône and Po towards the Mediterranean). An integration of maritime harbours will notably enlighten the hubs between maritime and riverine transport zones. Those hubs are at the margin of our current selective network model, but in the very centre of real life networks. For comparison, we resorted to a similar weighted model of transport on land, river and sea for the entire Imperium Romanum, the "ORBIS Stanford Geospatial Network Model of the Roman World", created by Walter Scheidel and his colleague over the last years to estimate costs of transport and the spatial integration of the Empire. Also ORBIS works on the basis of a transport-cost-weighted network of (terrestrial, riverine and maritime) routes between places.[22] As it targets the entire Empire, it is of course less detailed and integrates a smaller number of places for Rhine and Danube as does our model, for instance. Still, the entire graph (**Fig. 19**) consists of 678 nodes and 1104 links (or 2209 directed arcs) and to our knowledge is the most sophisticated network model of transport and traffic in the Roman World we have to date. Fortunately, the data is open access available online, so we could re-create the ORBIS-network (with some modifications) in order to execute an analysis for the purpose of comparison with a focus on the Danube/Rhine/Rhône/Po-areas. If we determine the betweenness-centrality of the nodes in the ORBIS-network (**Fig. 20**), differences to our restricted model are significant: the centrality values are much more unequally distributed and focused on hubs connecting different regions of the empire and especially providing connectivity to the maritime routes.

# Case Studies Byzantium

Within the Priority-Programme 1630, the project "Harbours and landing places on the Balkan coasts of the Byzantine Empire (4th to 12th centuries)" aims at a systematic survey of all relevant maritime sites on the basis of historical and archaeological evidence.[23] Based on the data collected for the coasts of modern-day North-west Greece, we modelled another series of transport networks. Since routes on the open sea are not determined in the way riverine connections are, we modelled a "Maximum Distance Network", in which nodes (harbour places) are linked to all other nodes within a set (geographic) distance. In order "to reflect logistics of travelling", we chose a cut-off distance of 125 km, which would have been a distance travelled by a ship on one day under very good conditions.[24] Furthermore, in order to integrate transport costs, links are again weighted by using the inverted geographical distance between them; for future studies, recent tools in the GIS-based calculation of cost surfaces for maritime travel could be applied to achieve more appropriate results in this regard.[25]

---

[21] The distribution map is based on Grunwald 2012. Cf. Gross 2012;

[22] Scheidel/Meeks et al.: http://orbis.stanford.edu/. The ORBIS-data set can be downloaded via the following link: https://purl.stanford.edu/mn425tz9757 (Creative Commons Attribution 3.0 Unported License); authors of the data set are E. Meeks/W.Scheidel/J.Weiland/S. Arcenas. For the reconstruction of searoutes, the creators of ORBIS especially used Arnaud 2005.

We created such network models for seven time slices (6[th], 7[th], 8[th], 9[th], 10[th], 11[th] and 12[th] cent. CE) on the basis of the respective data for the use of harbour sites and determined the various measures on the node- and network-level **(Fig. 21 and 22)**. As the visualisations and the measures for the seven time slices show, similar to some of the river network models, we observe a reduction of the scale of the network (with regard to the number of nodes and links) from the 6[th] to the 7[th] and 8[th] century, before the network "grows" again from the 9[th] century onwards **(Fig. 23)**. At the same time, the decrease is less dramatic with regard to measures such as density, clustering coefficient and transitivity, which would indicate that – on a smaller scale – maritime connectivity was still maintained across the region **(Fig. 24)**; only for the 9[th] century, the network is fragmented in two components, but this may also result from a lack of data **(Fig. 23)**.[26]

On the level of individual nodes, results on the (in many cases significantly) changing relative position of a harbour within the series of models for the seven centuries can be compared with our knowledge on the development of sites from archaeological and historical evidence. In the case of Naupaktos, for instance, the increase in the betweenness centrality of the node correlates with the emergence of the city as Byzantine centre both of traffic and of administration in Western Greece from the 9[th] century onwards. The same is true for the harbours of Patrai on the Peloponnese and for the newly emerging port of Bonditza in the Ambrakian Gulf **(Fig. 25)**.[27] Our network models thus can contribute to an understanding of the interplay between a site´s relative position within the possible web of maritime routes and its urban development.

---

[23] http://www.spp-haefen.de/en/projects/byzantine-harbours-on-the-balkan-coasts/.
[24] Cf. Mol/Hoogland/Hofman 2015, 275-305, esp. 281-283. For the cut-off distance cf. also Preiser-Kapeller 2013.
[25] Leidwanger 2013.
[26] Cf. Veikou 2012.
[27] Cf. Heher/Preiser-Kapeller/Simeonov 2015.

Recently, Thomas Tartaron has highlighted the significance of short-distance connections often overlooked in the favour of long-distance networks (in his case, for the Bronze Age Aegean): "*More importantly, such long-distance connections were dwarfed in quantity by dense networks of local and regional maritime connections among (...) communities. (...) they must have dominated the use of anchorages, large and small (...). Local and microregional maritime networks are best expressed by the concept of the "small world" (...), composed of communities bound together by intensive, habitual interactions due to geography, traditional kinship ties, or other factors. (...) Small worlds are nested within larger regional and interregional economic and sometimes political networks.*"[28] Spatial proximity thus would be an essential parameter to estimate the actual "strength" (indicating the frequency or scale of exchange between sites, for instance) of connections.[29] Yet as analytical results by Ducruet et alii for maritime traffic in the modern period or by Sindbæk for the early medieval North Sea indicate, clusters do not only necessarily emerge among nodes spatially near to each other. Especially ties of long distance exchange could connect ports more densely with other hubs of commerce oversea than with nearby sites of less mercantile activity. Similarly, measures of superior political or religious authorities or the desire to connect to sacred places etc. could "rewire" a considerable amount of the links of localities to more far away sites.[30]

Distribution patterns of artefacts hint at such different "logics of connectivity"; distribution patterns can also be turned into networks as "affiliation" models. In such a model, first sites and various types of artefacts are connected with each other. This "2mode-network" (with two categories of nodes) is transformed into a "1mode-network", in which two sites are connected with each other if at least one type of artefact can be found in both of them **(Fig. 26)**. Links in such a network thus are "ties of similarity" between artefact assemblages in sites and not ties of direct interaction or exchange (a fact sometimes overlooked, especially if such networks are visualised on a geographic map). Through the integration of several distributed types of artefacts these "ties of similarity" are weighted; centrality measures for individual nodes indicate different degrees of integration in what we can assume were systems of distribution and exchange. As Søren Sindbæk had outlined, we gain some insights into difference between "hubs" and peripheries within the distribution system and may be able to identify "clusters of similarities" indicating more intense exchange among these sites.[31]

---

[28] Tartaron 201, 6-7, 80, 88.
[29] Isaksen 2008; Barthélemy 2011; Gorenflo/Bell 1991.
[30] Ducruet/Zaidi 2012; Malkin 2011; Sindbæk 2007; Sindbæk 2013. For this aspect in the Early Medieval West using the example of monastic networks of exchange cf. Haase/Werther/Wunschel 2015, 151–189.
[31] Sindbæk 2007; Sindbæk 2013.

We created two such models for the same region – 20 sites in the late medieval Peloponnese in the 13th-15th century, one based on the distribution of nine types of locally produced ceramics and one based on the distribution of 14 types of imported ceramics (from places across the entire Mediterranean). As analysed in detail elsewhere, both emerging 1mode-networks between sites show a considerable amount of overlap, but also significantly different focal points of distribution as expressed in the degree centrality (= the accumulated strength of links of one node) of sites, which only very weakly correlates between the two models **(Fig. 27 and 28)**. But both network models have in common that the strength of these ties of similarity of artefact assemblages does not follow the "logic of spatial proximity", as an analysis of the correlation between distances between nodes and the strength of ties between them indicates **(Fig. 29 and 30)**. On the contrast, we have to reckon with hubs of distribution more similar to each other across distances than with less important places probably nearby **(Fig. 31)**.[32] Yet, again the "logic of spatial proximity" may help us to unlock further underlying dynamics: if we overlay a maximum distance network on the affiliation networks and cut off all ties beyond a distance of 50 km, we detect again distinct local clusters around central places – connected among themselves through over-regional ties – and surrounding sites of lesser centrality, which may have been served by the central market places (such as Corinth) **(Fig. 32 and 33)**. In this particular case we can also compare the results of our analysis with written evidence for actual commercial contacts in that period, which confirm the model´s explanatory value[33]; this indicates its applicability to other data sets where this is not the case.

# Conclusion

As our "relational approach" indicates, harbours and ports can be understood as emerging "hubs of connectivity" in traffic systems; tools of network analysis can help us to capture, visualise and analyse aspects of these systems and the interplay between routes, the distribution of artefacts or the mobility of individuals and the emergence and dynamics of individual sites. An overlap of such different data sets contributes to an even better understanding of the actual complexity of webs of exchange and interaction across the sea or via rivers and lakes and allows for assumptions on the character and limits of local, regional and supra-regional systems of traffic and commerce. On this basis, also a comparison of structural properties of harbour networks across regions and periods becomes possible – one central aim of the Priority-Programme 1630. In a next step, we also aim to combine network models for the river systems of Northern Italy and Central Europe with those for the Late Roman/Byzantine Balkans, thus contributing to a truly "European" approach towards traffic systems of the past. Taking into account these potentials, we hope for a further integration of these tools and concepts into harbour research.

---

[32] Preiser-Kapeller 2014. For the data see Vroom 2011.
[33] Vroom 2011; Jacoby 2013.

# Tables

| | Po Period I | Po Period II | Rhône Period I | Rhône Period II | Rhine Period I | Rhine Period II | Danube Period I | Danube Period II | Connected Period I |
|---|---|---|---|---|---|---|---|---|---|
| **Number of nodes** | 22 | 17 | 31 | 22 | 59 | 55 | 15 | 13 | 128 |
| **Number of edges** | 36 | 19 | 72 | 21 | 124 | 120 | 17 | 16 | 342 |
| **Density** | 0,156 | 0,136 | 0,077 | 0,091 | 0,036 | 0,04 | 0,162 | 0,205 | 0,021 |
| **Clustering Coefficient** | 0,505 | 0,265 | 0,273 | 0 | 0,136 | 0,154 | 0,233 | 0,308 | 0,234 |
| **Diameter** | 13 | 25 | 56 | 93 | 207 | 261 | 23 | 16 | 183 |
| **Betweenness Centralisation** | 0,435 | 0,436 | 0,502 | 0,295 | 0,295 | 0,425 | 0,347 | 0,359 | 0,435 |
| **Transitivity** | 0,657 | 0,250 | 0,341 | 0 | 0,191 | 0,242 | 0,462 | 0,455 | 0,346 |
| **Circuitry (Alpha-index)** | 0,385 | 0,103 | 0,74 | 0,00 | 0,58 | 0,63 | 0,12 | 0,19 | 0,86 |

**Tab. 1:** Comparison of network measures for the network models of riverine transport in period I (1[st]-5[th] cent. CE) and period II (6[th]– early 11[th] cent. CE).

# Figures

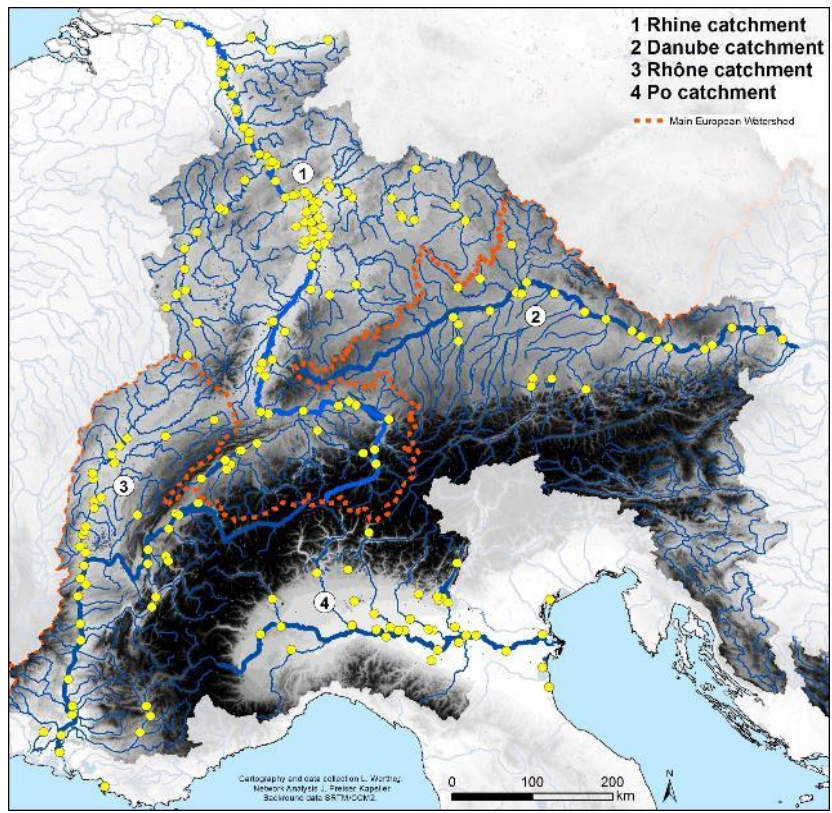

**Fig. 1** Harbours and landing sites integrated in the network models of riverine transport in the catchments of Danube, Po, Rhine and Rhône. – (Cartography and Data Collection L. Werther, Universität Jena).

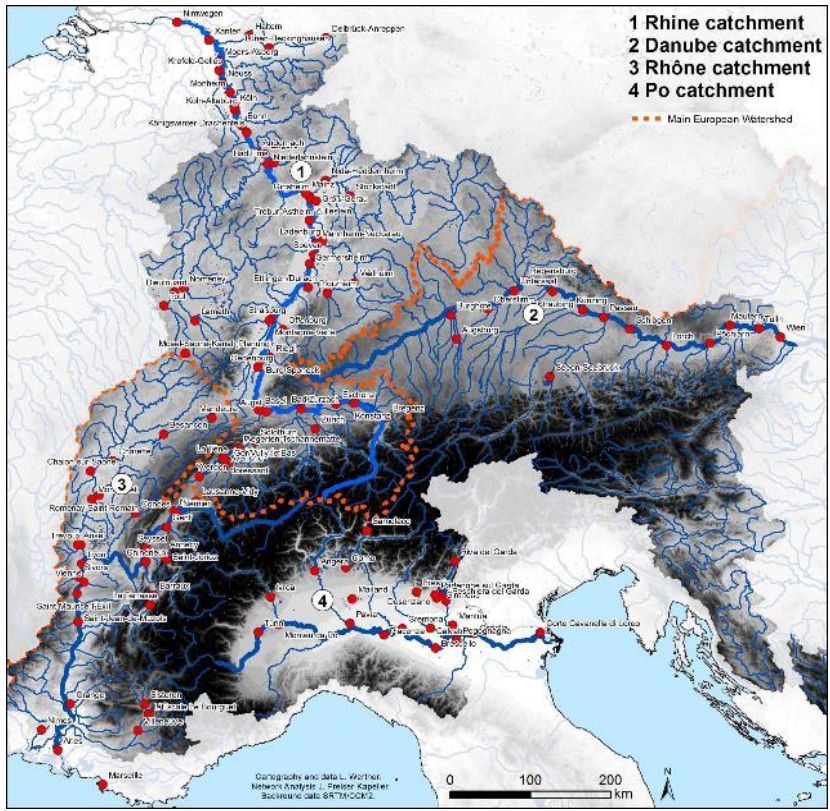

**Fig. 2** Harbours and landing sites integrated in the network models of riverine transport in the catchments of Danube, Po, Rhine and Rhône for period I(1$^{st}$-5$^{th}$ cent. CE). – (Cartography and Data Collection L. Werther, Universität Jena).

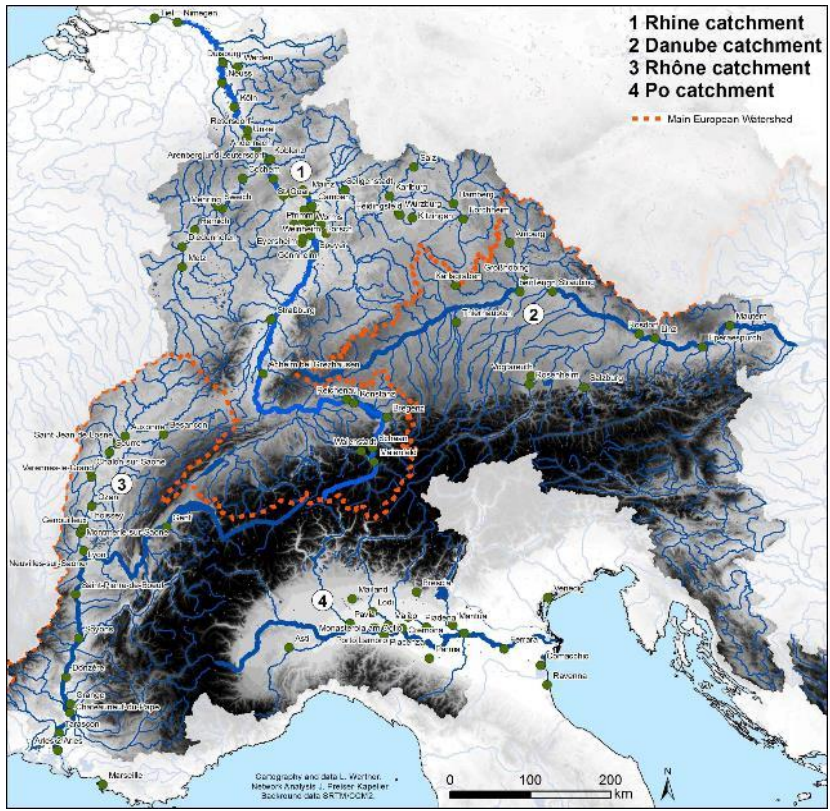

**Fig. 3** Harbours and landing sites integrated in the network models of riverine transport in the catchments of Danube, Po, Rhine and Rhône for period II (6th– early 11th cent. CE). – (Cartography and Data Collection L. Werther, Universität Jena).

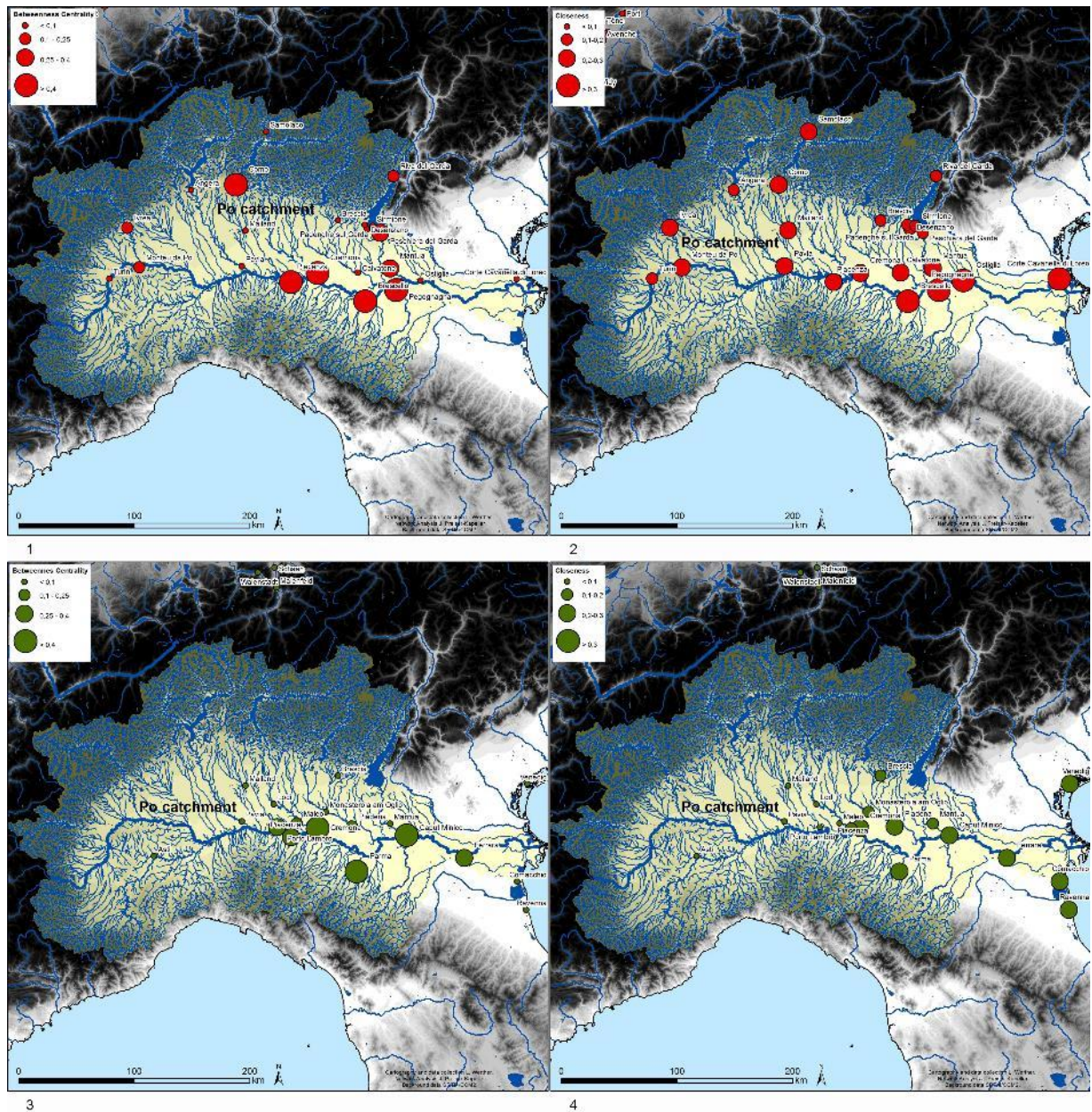

**Fig. 4** Nodes in the network model of riverine transport in the Po catchment: **1** In period I (1st-5th cent. CE) sized according to their betweenness centrality. – **2** In period I (1st-5th cent. CE) sized according to their closeness centrality. – **3** In period II (6th– early 11th cent. CE) sized according to their betweenness centrality. – **4**. In period II (6th– early 11th cent. CE) sized according to their closeness centrality. – (Cartography and Data Collection L. Werther, Universität Jena; Network Analysis J. Preiser-Kapeller, ÖAW).

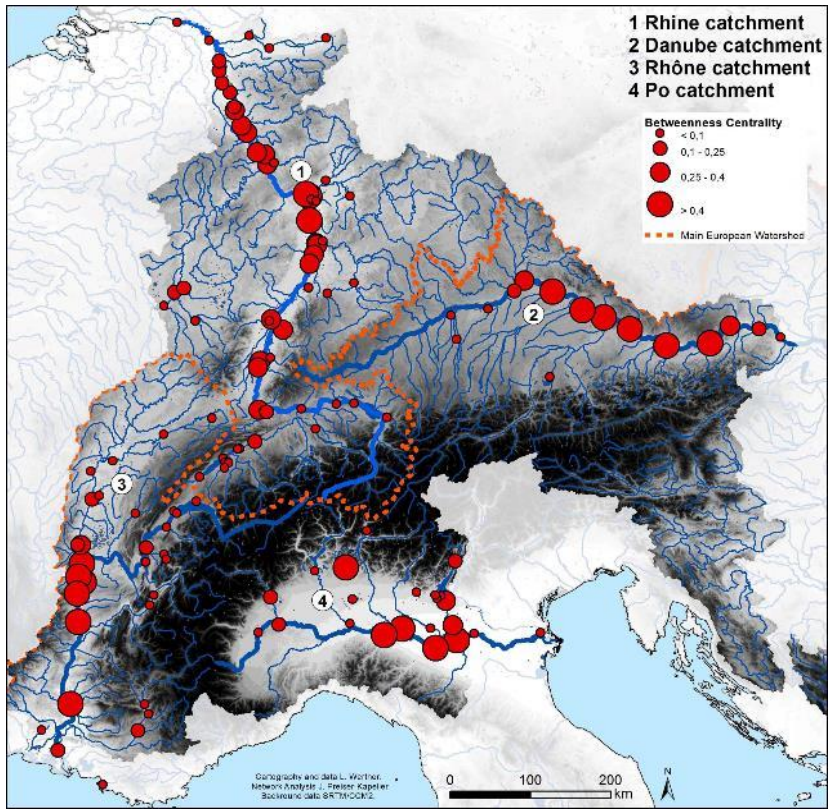

**Fig. 5** Nodes in the network model of riverine transport in the catchments of Danube, Po, Rhine and Rhône for period I (1ˢᵗ-5ᵗʰ cent. CE) sized according to their betweenness centrality. – (Cartography and Data Collection L. Werther, Universität Jena; Network Analysis J. Preiser-Kapeller, ÖAW).

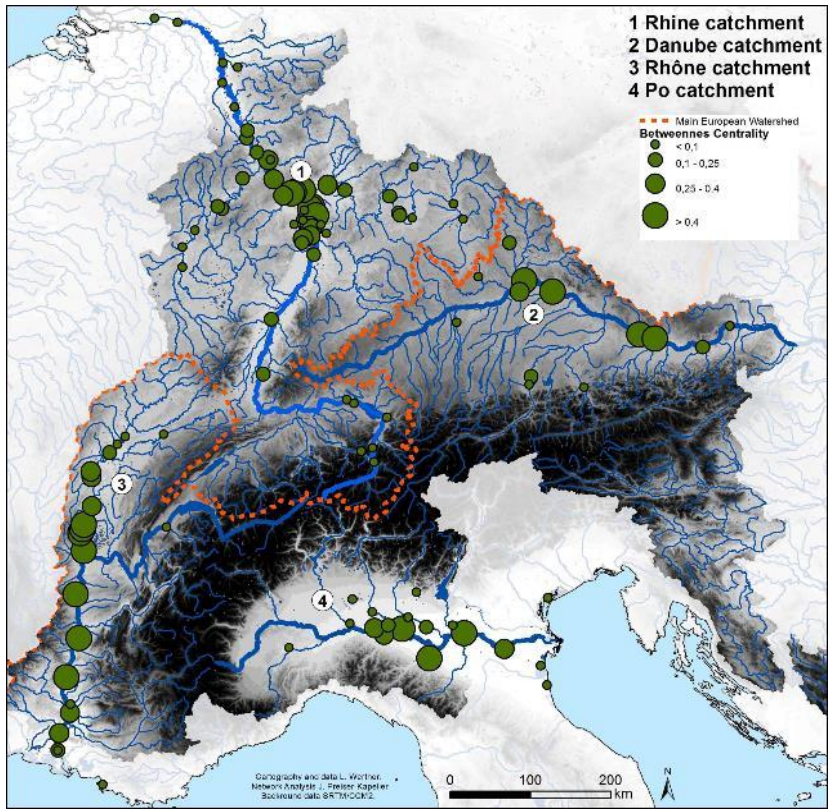

**Fig. 6** Nodes in the network model of riverine transport in the catchments of Danube, Po, Rhine and Rhône for period II (6th– early 11th cent. CE) sized according to their betweenness centrality. – (Cartography and Data Collection L. Werther, Universität Jena; Network Analysis J. Preiser-Kapeller, ÖAW).

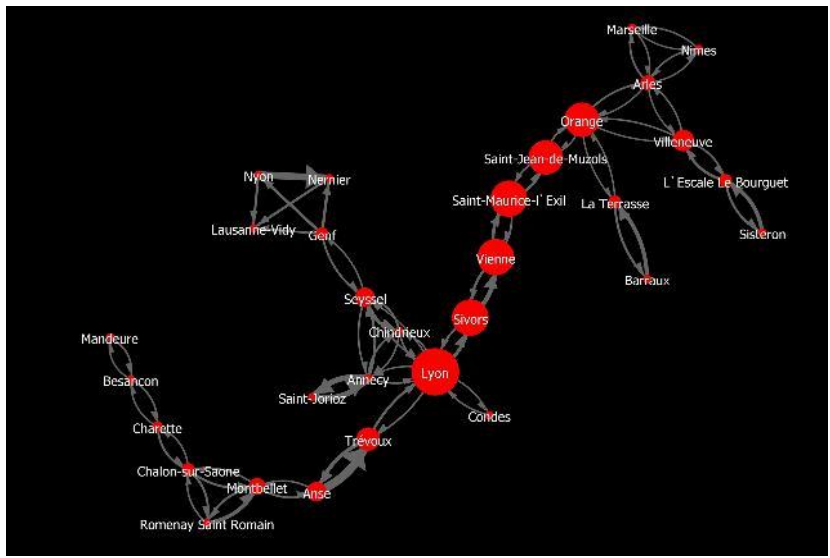

**Fig. 7** Topological graph of the weighted and directed network model of riverine transport in period I (1st-5th cent. CE) in the Rhône catchment. Nodes are sized according to their betweenness centrality. . – (Data Collection L. Werther, Universität Jena; Network Analysis J. Preiser-Kapeller, ÖAW).

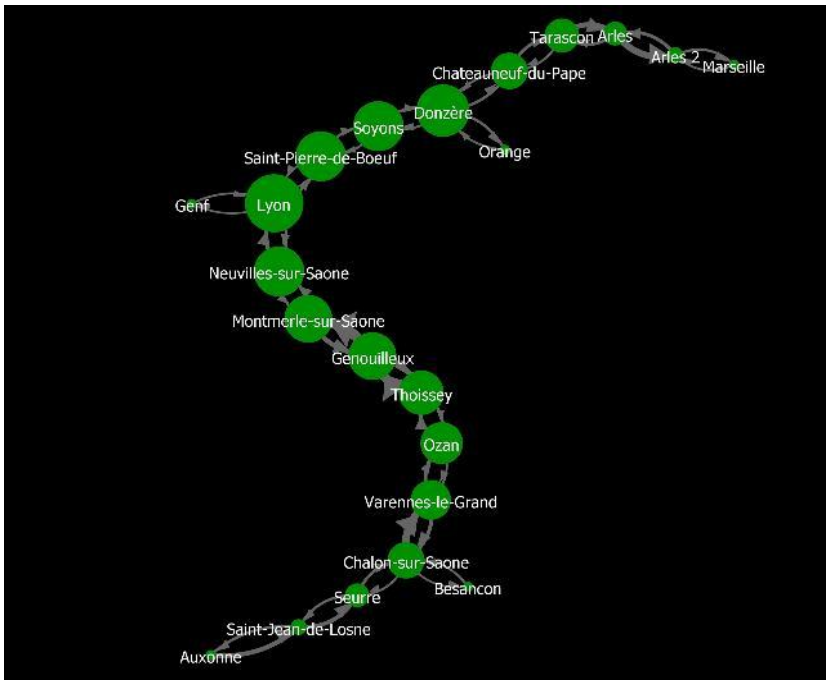

**Fig. 8** Topological graph of the weighted and directed network model of riverine transport in period II (6[th]– early 11[th] cent. CE) in the Rhône catchment. Nodes are sized according to their betweenness centrality. – (Data Collection L. Werther, Universität Jena; Network Analysis J. Preiser-Kapeller, ÖAW).

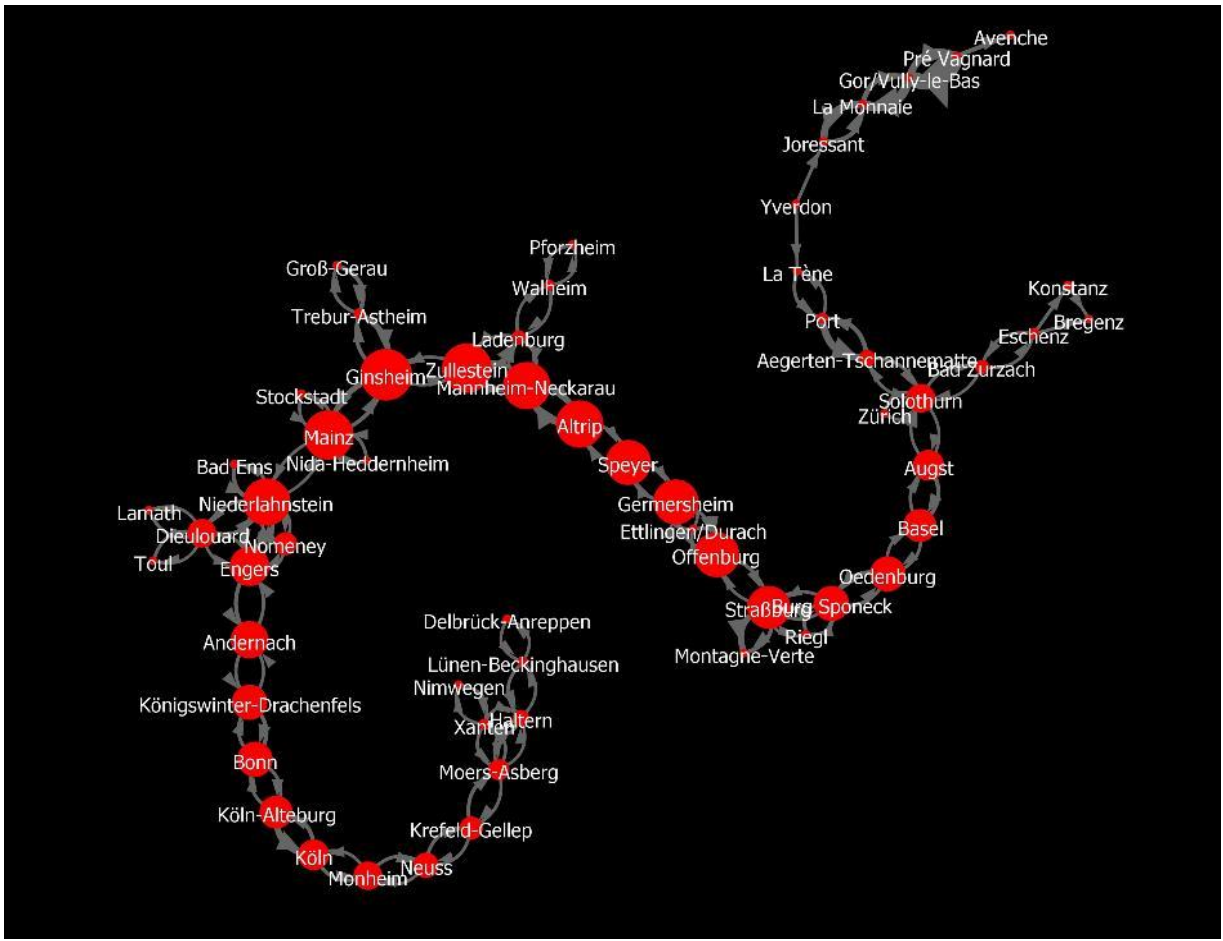

**Fig. 9** Topological graph of the weighted and directed network model of riverine transport in period I (1st-5th cent. CE) in the Rhine catchment. Nodes are sized according to their betweenness centrality. – (Data Collection L. Werther, Universität Jena; Network Analysis J. Preiser-Kapeller, ÖAW).

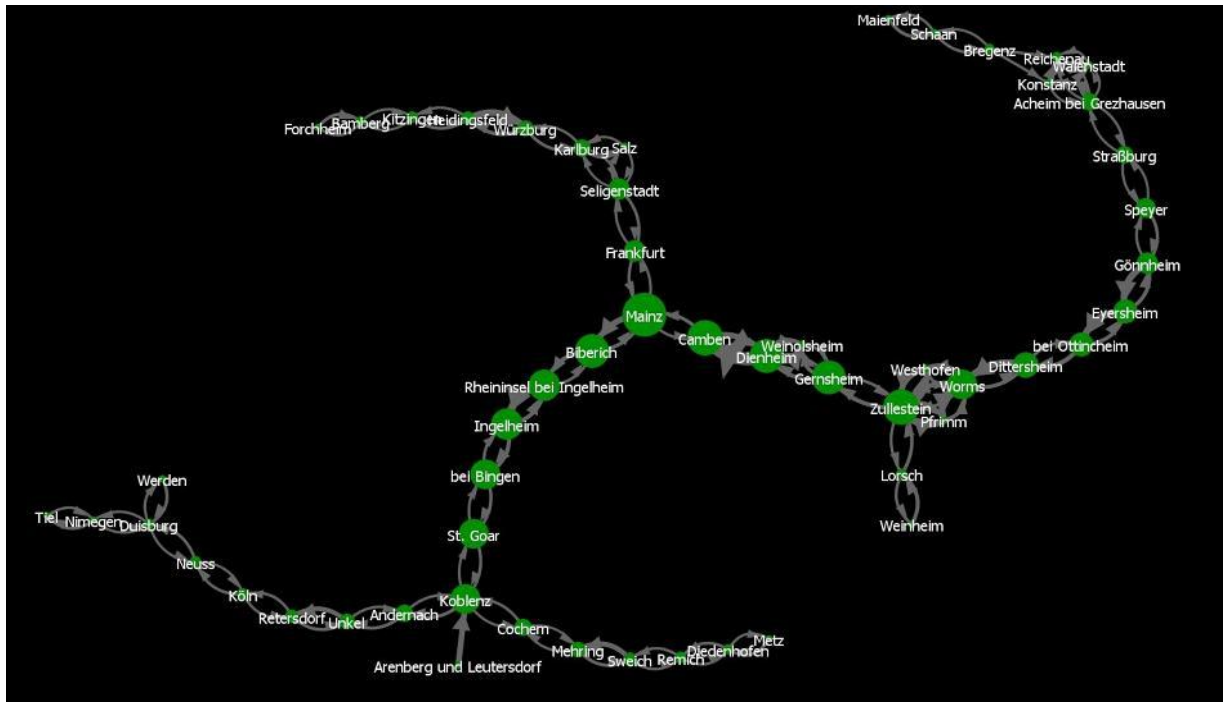

**Fig. 10** Topological graph of the weighted and directed network model of riverine transport in period II (6[th]– early 11[th] cent. CE) in the Rhine catchment. Nodes are sized according to their betweenness centrality. – (Data Collection L. Werther, Universität Jena; Network Analysis J. Preiser-Kapeller, ÖAW).

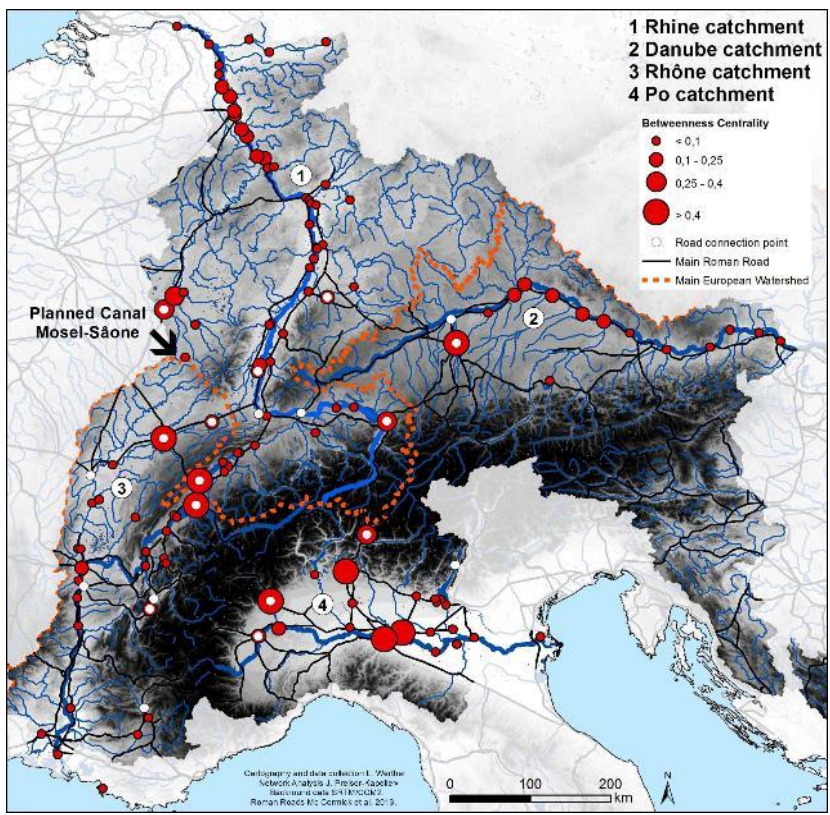

**Fig. 11** Nodes in the interconnected network model of riverine transport in the catchments of Danube, Po, Rhine and Rhône for period I (1[st]-5[th] cent. CE) sized according to their betweenness centrality. – (Cartography and Data Collection L. Werther, Universität Jena; Network Analysis J. Preiser-Kapeller, ÖAW).

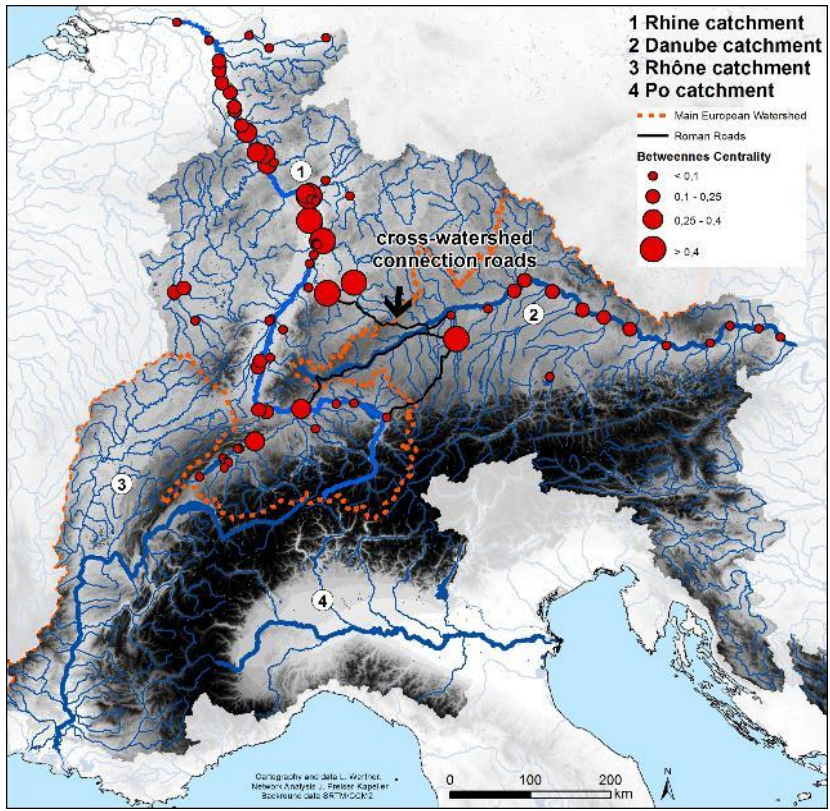

**Fig. 12** Nodes in the interconnected network model of riverine transport in the catchments of Danube and Rhine for period I (1st-5th cent. CE) sized according to their betweenness centrality. – (Cartography and Data Collection L. Werther, Universität Jena; Network Analysis J. Preiser-Kapeller, ÖAW).

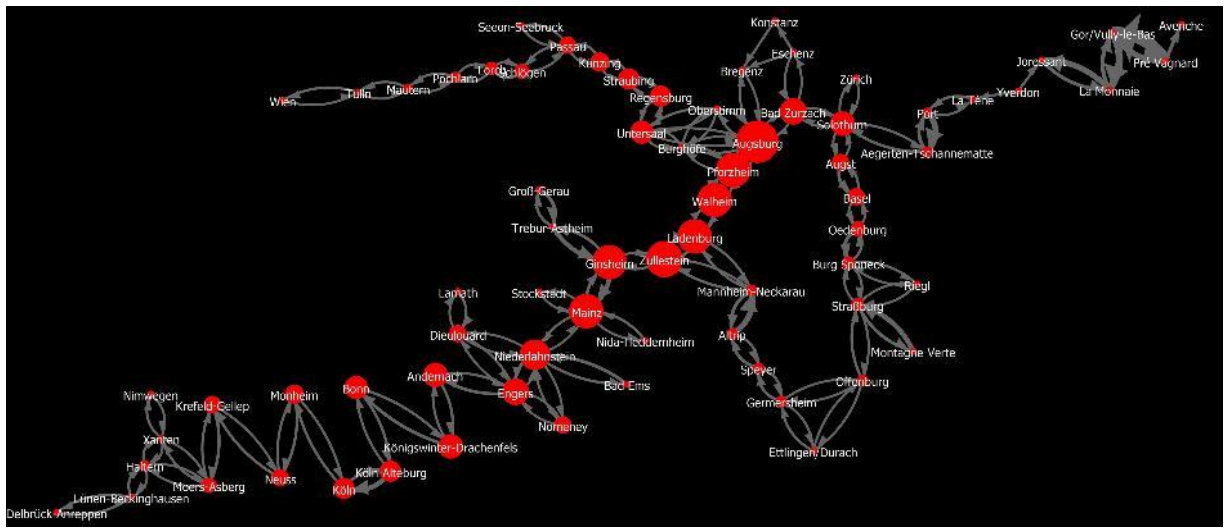

**Fig. 13** Topological graph of the weighted and directed interconnected network model of riverine transport in the catchments of Danube and Rhine in period I (1st-5th cent. CE). Nodes are sized according to their betweenness centrality. – (Data Collection L. Werther, Universität Jena; Network Analysis J. Preiser-Kapeller, ÖAW).

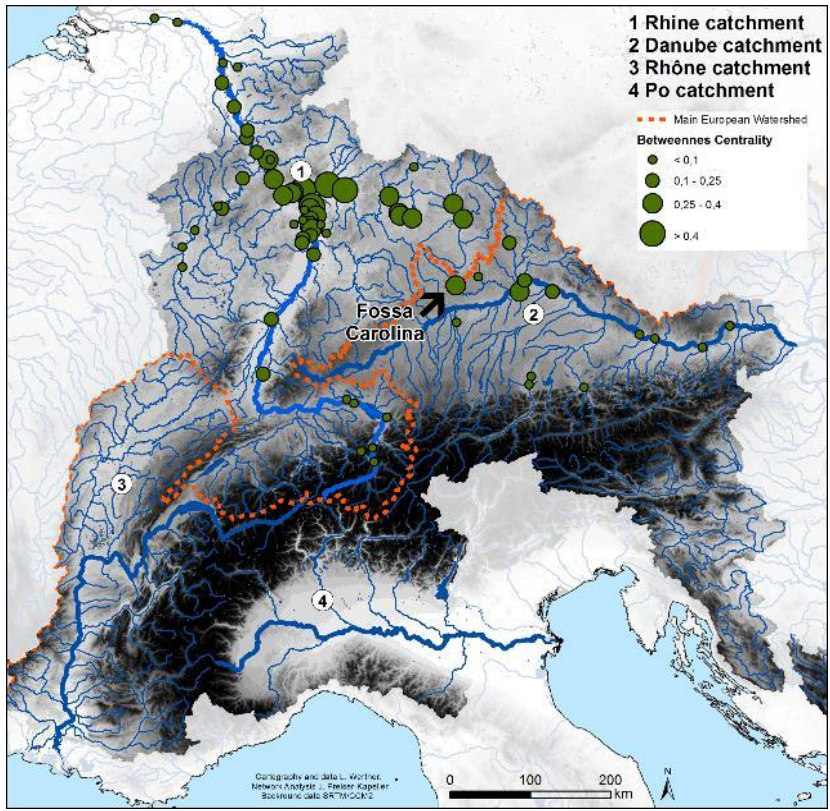

**Fig. 14** Nodes in the interconnected network model of riverine transport in the catchments of Danube and Rhine with the Fossa Carolina for period II (6<sup>th</sup>– early 11<sup>th</sup> cent. CE) sized according to their betweenness centrality. – (Cartography and Data Collection L. Werther, Universität Jena; Network Analysis J. Preiser-Kapeller, ÖAW).

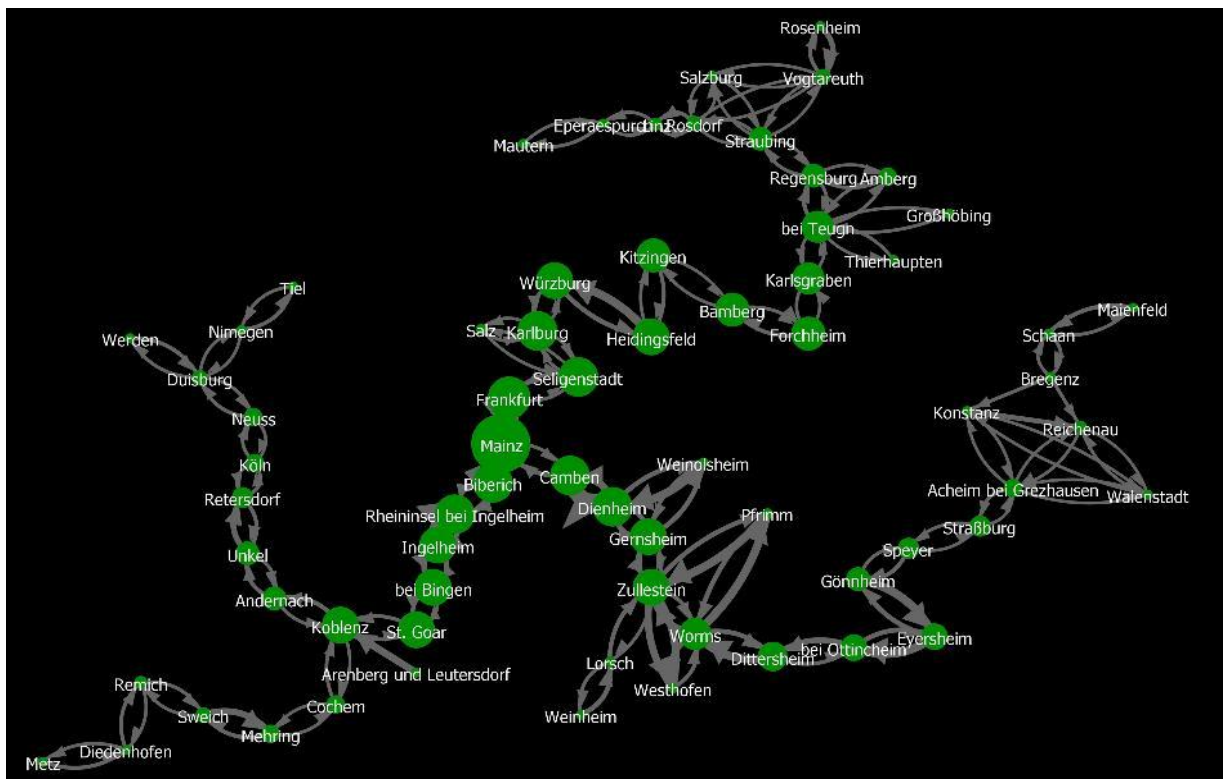

**Fig. 15** Topological graph of the weighted and directed interconnected network model of riverine transport in the catchments of Danube and Rhine with the Fossa Carolina in period II (6<sup>th</sup>– early 11<sup>th</sup>

cent. CE). Nodes are sized according to their betweenness centrality. – (Data Collection L. Werther, Universität Jena; Network Analysis J. Preiser-Kapeller, ÖAW).

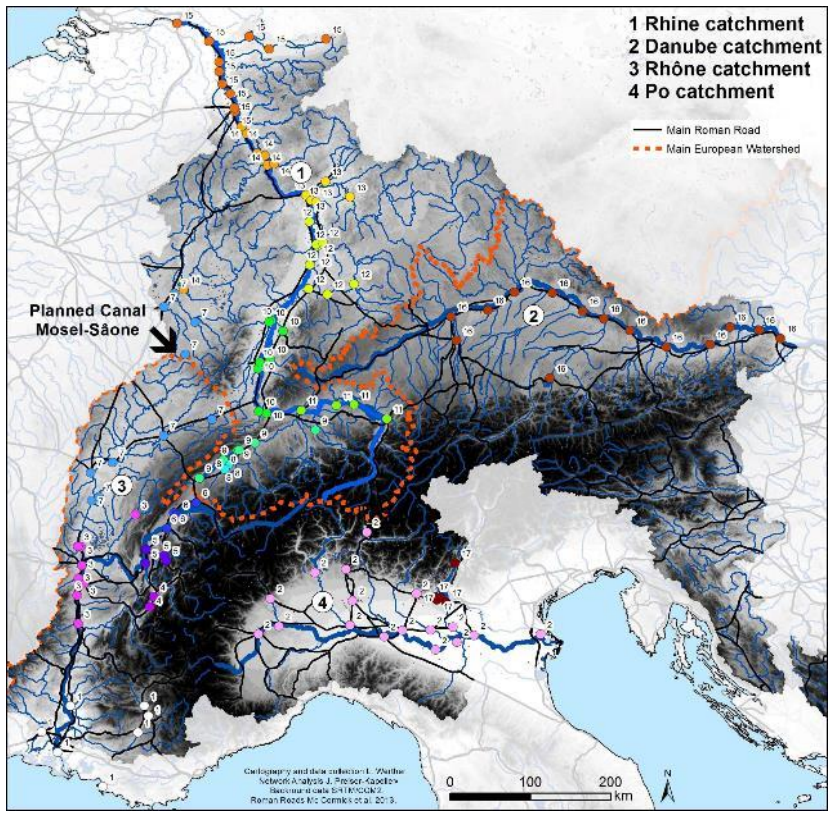

**Fig. 16** Clusters identified in the in the interconnected network model of riverine transport in the catchments of Danube, Po, Rhine and Rhône for period I (1st-5th cent. CE) with the help of the Newman-algorithm. – (Cartography and Data Collection L. Werther, Universität Jena; Network Analysis J. Preiser-Kapeller, ÖAW).

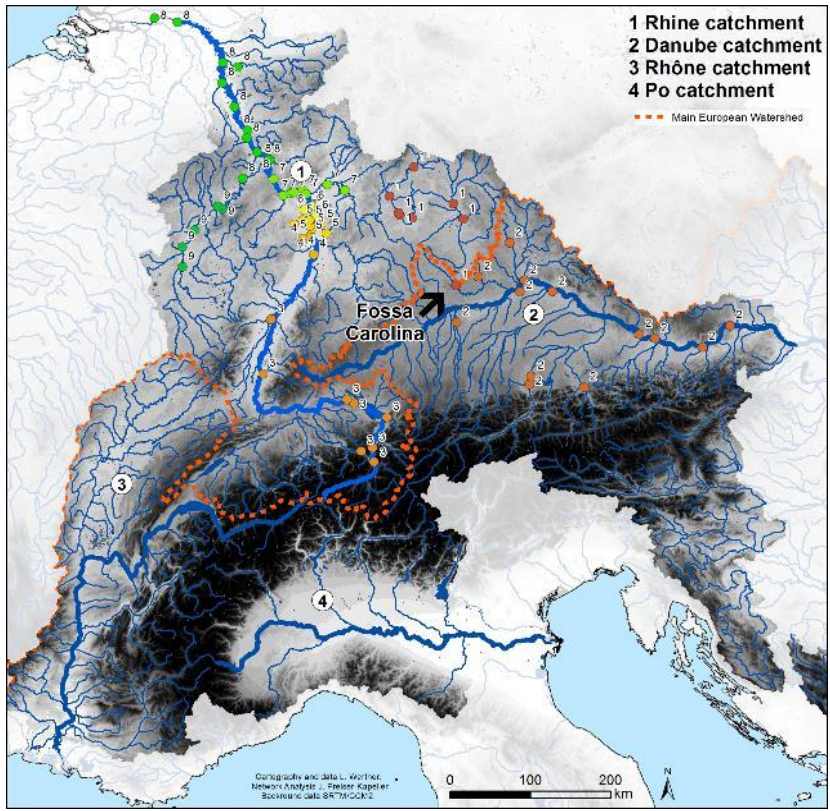

**Fig. 17** Clusters identified in the in the interconnected network model of riverine transport in the catchments of Danube and Rhine with the Fossa Carolina in period II (6th– early 11th cent. CE) with the help of the Newman-algorithm. – (Cartography and Data Collection L. Werther, Universität Jena; Network Analysis J. Preiser-Kapeller, ÖAW).

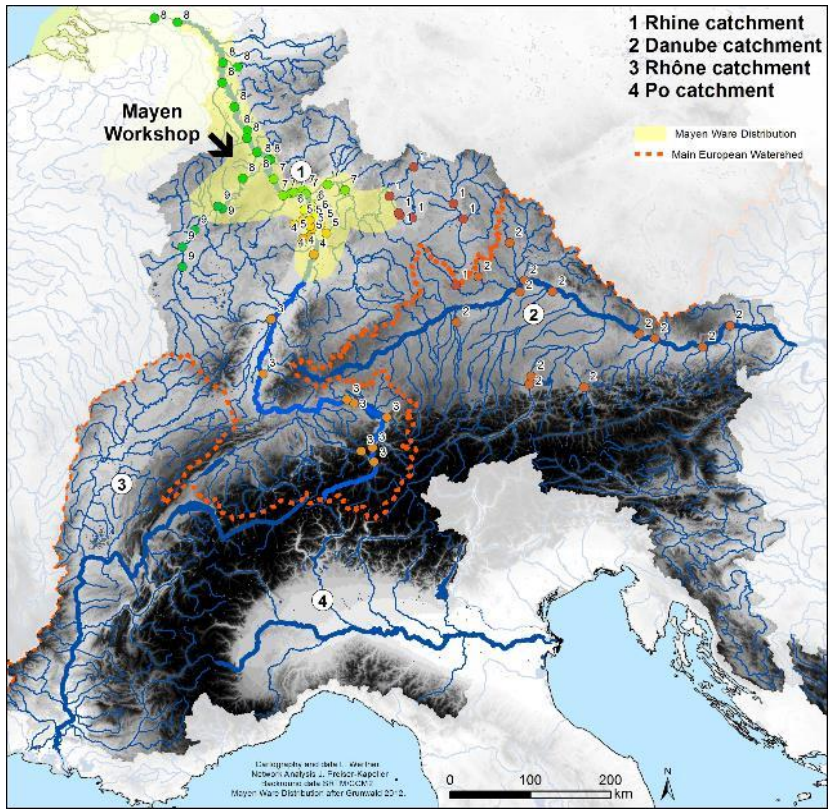

**Fig. 18** Clusters identified in the in the interconnected network model of riverine transport in the catchments of Danube and Rhine with the Fossa Carolina in period II (6[th]– early 11[th] cent. CE) with the help of the Newman-algorithm, combined with the distribution of Early and High Medieval "Mayener Ware" after Grunwald 2012. – (Cartography and Data Collection L. Werther, Universität Jena; Network Analysis J. Preiser-Kapeller, ÖAW).

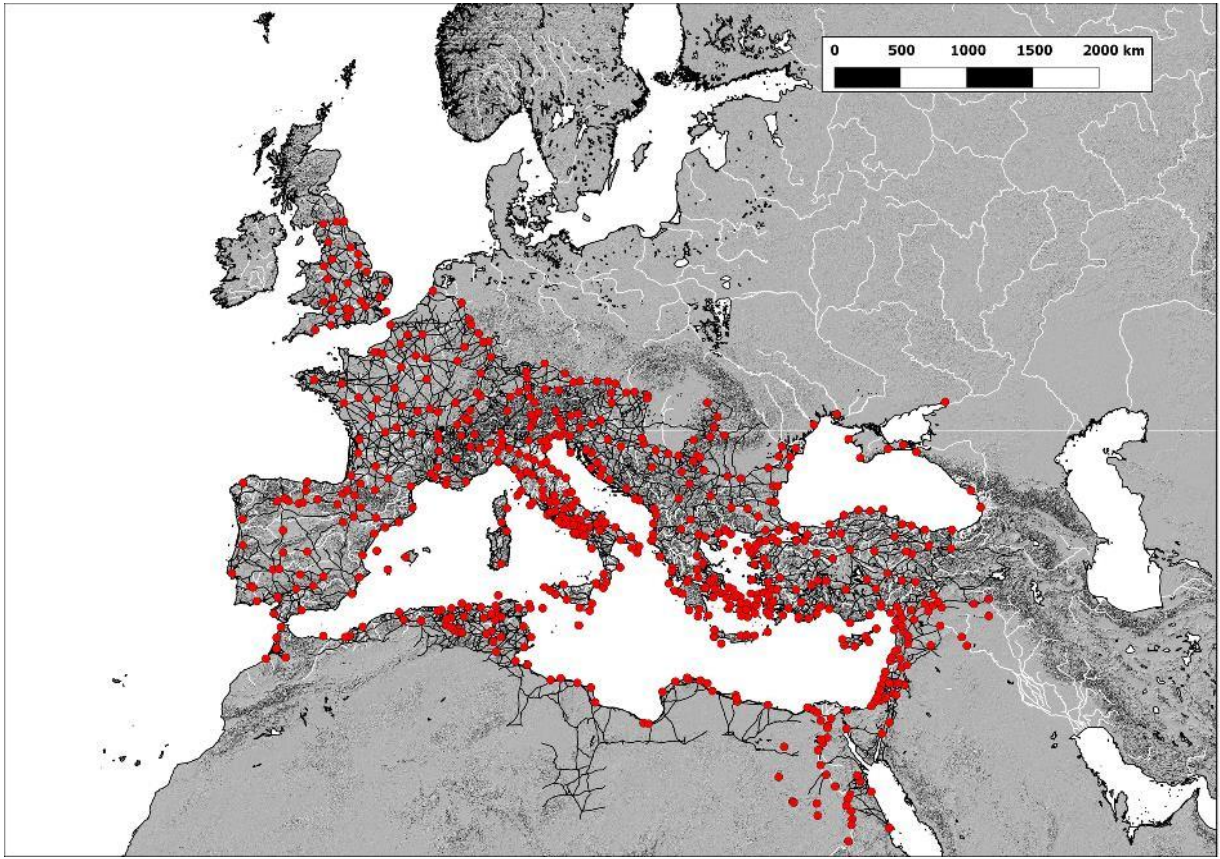

**Fig. 19** Nodes in the "ORBIS Stanford Geospatial Network Model of the Roman World". – (J. Preiser-Kapeller, ÖAW).

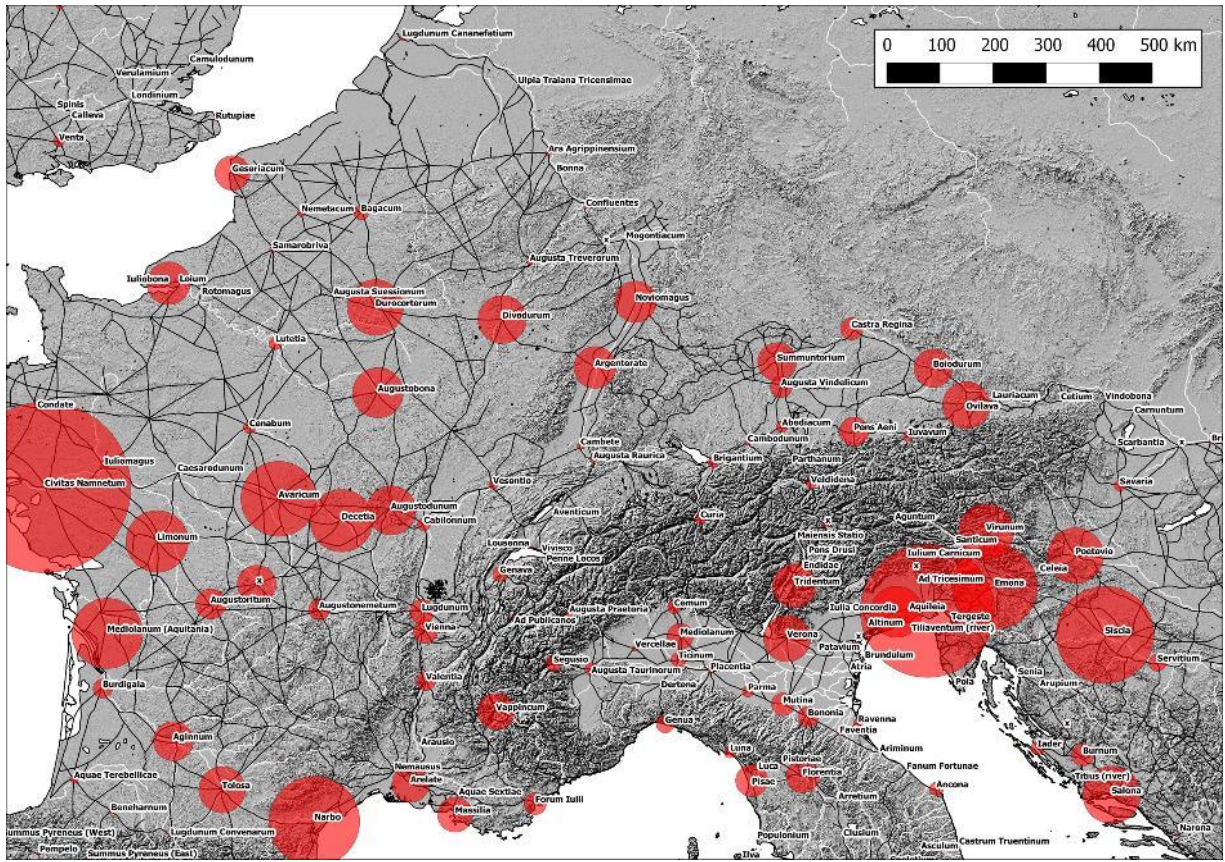

**Fig. 20** Nodes in the "ORBIS Stanford Geospatial Network Model of the Roman World" sized according to their betweenness centrality. – (J. Preiser-Kapeller, ÖAW).

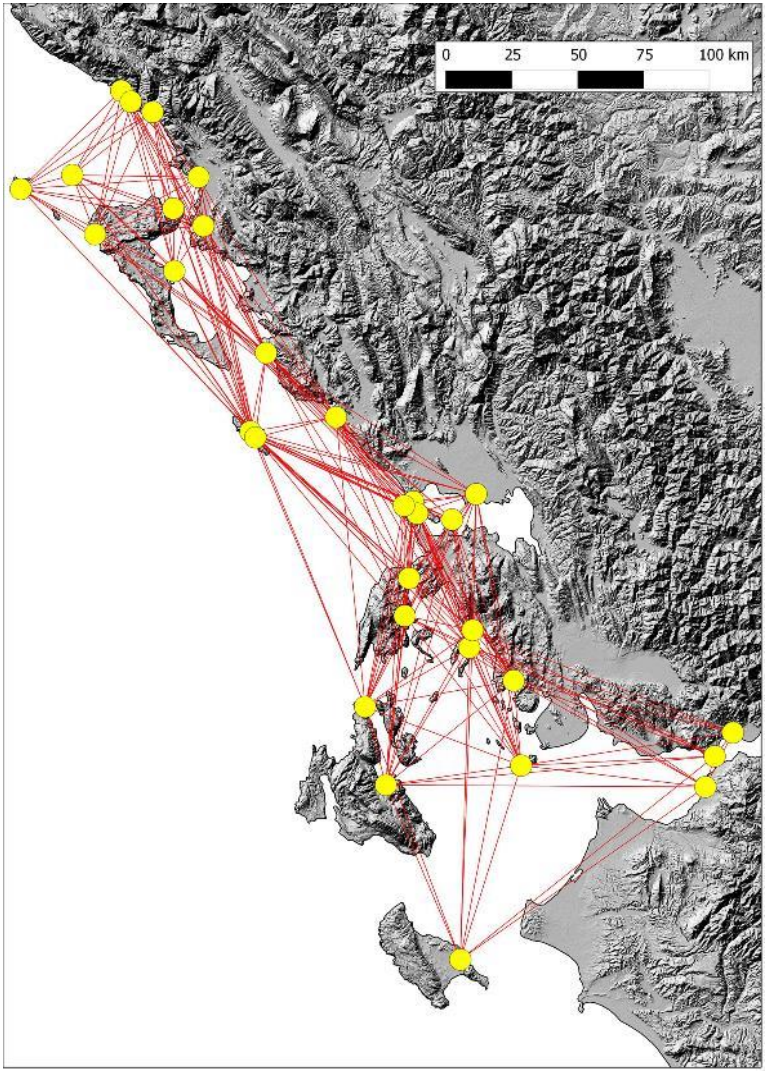

**Fig. 21** Weighted maximum distance network model for harbour sites in Western Greece, 6[th] cent. CE (Cartography, data collection and network analysis J. Preiser-Kapeller, ÖAW).

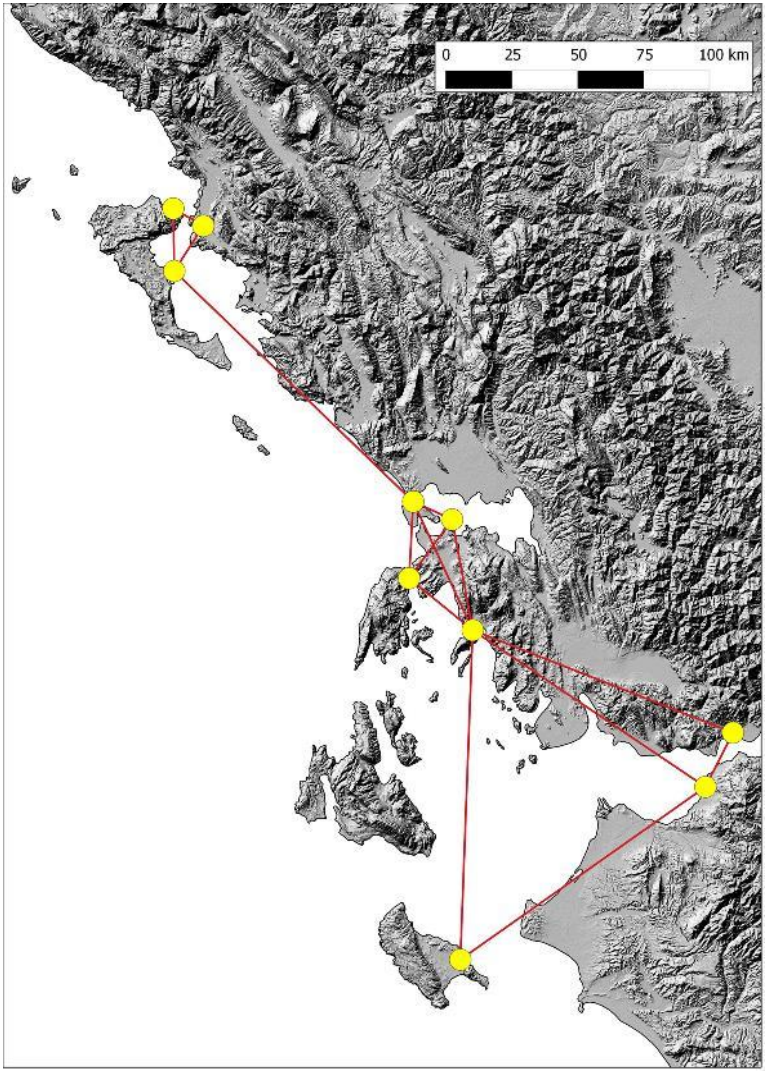

**Fig. 22** Weighted maximum distance network model for harbour sites in Western Greece, 8th cent. CE (Cartography, data collection and network analysis J. Preiser-Kapeller, ÖAW).

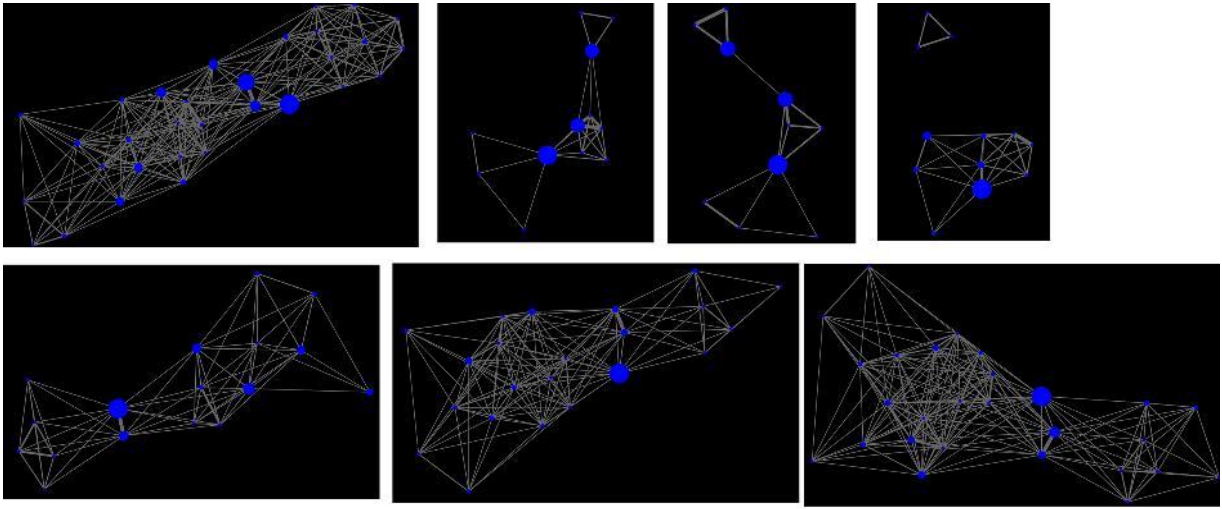

**Fig. 23** Weighted maximum distance network model for harbour sites in Western Greece, 6th, 7th, 8th, 9th cent. CE (upper series of graphs), 10th, 11th, 12th cent. CE (lower series of graphs) (data collection and network analysis J. Preiser-Kapeller, ÖAW).

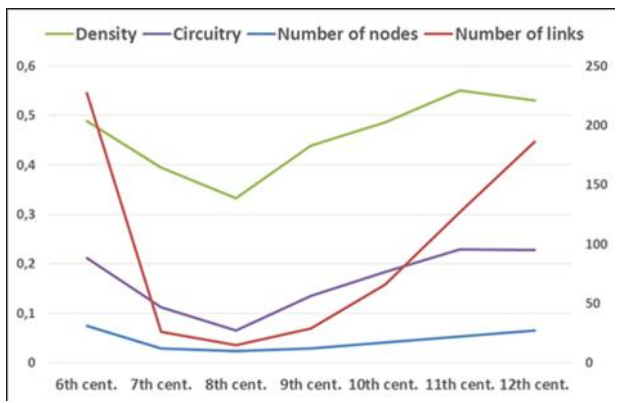

**Fig. 24** Network measures for all time slices of the weighted maximum distance network model for harbour sites in Western Greece, 6th-12th cent. CE (network analysis J. Preiser-Kapeller, ÖAW).

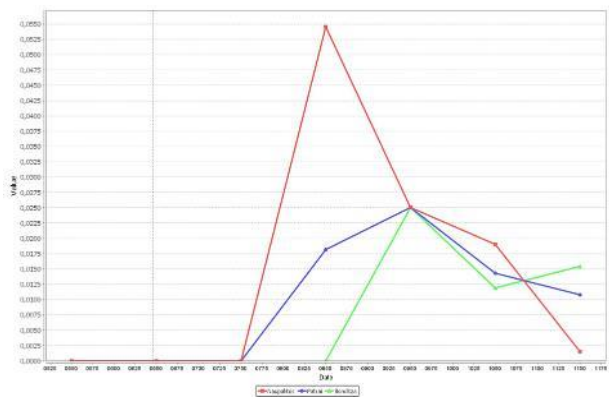

**Fig. 25** Betweenness centrality measures for three harbour sites in all time slices of the weighted maximum distance network model for harbour sites in Western Greece, 6th-12th cent. CE (network analysis J. Preiser-Kapeller, ÖAW).

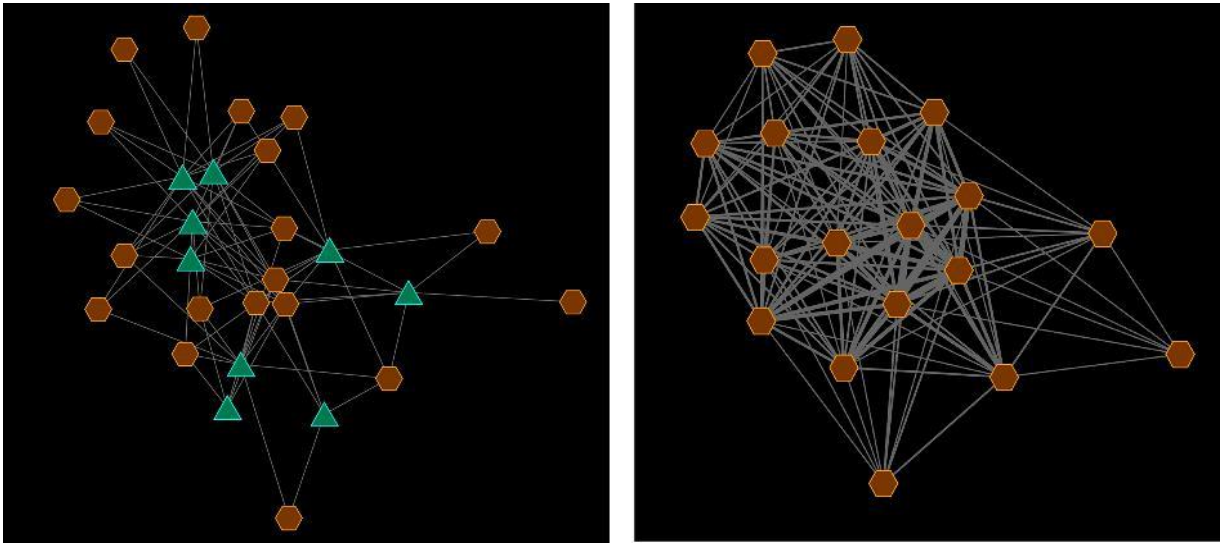

**Fig. 26** Left: affiliation network-model (2mode-network) with nine types of artefacts (triangles) and 19 sites (hexagons) where one or more of these types occurs; right: one-mode-network of sites based on the affiliation-network (two sites are connected on the basis of the co-occurrence of artefact types; the strength of links indicates the number of types common for two sites) (J. Preiser-Kapeller, ÖAW).

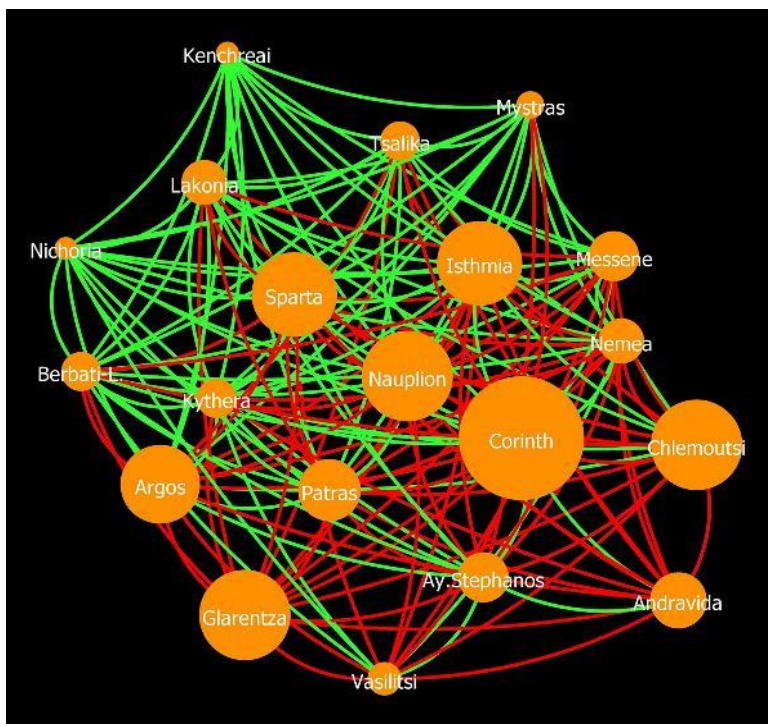

**Fig. 27** Network model of sites on the Peloponnese connected through the co-occurrence of types of locally produced ceramics (green) and of imported ceramics (red), 13th-15th cent. CE; nodes sized according to their degree centrality values in the network of imported ceramics (data Vroom 2011; graph and network analysis J. Preiser-Kapeller, ÖAW).

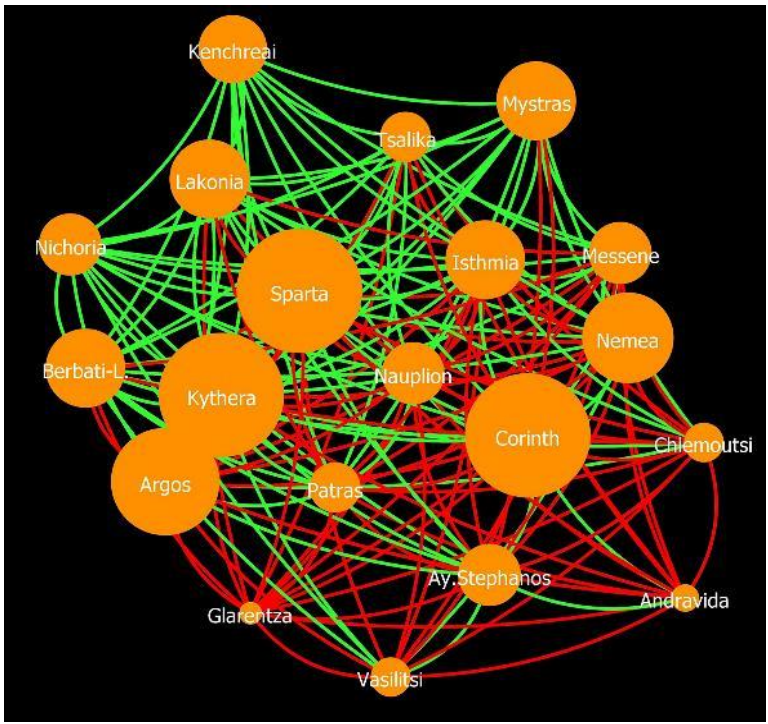

**Fig. 28** Network model of sites on the Peloponnese connected through the co-occurrence of types of locally produced ceramics (green) and of imported ceramics (red), 13th-15th cent. CE; nodes sized according to their degree centrality values in the network of locally produced ceramics (data Vroom 2011; graph and network analysis J. Preiser-Kapeller, ÖAW).

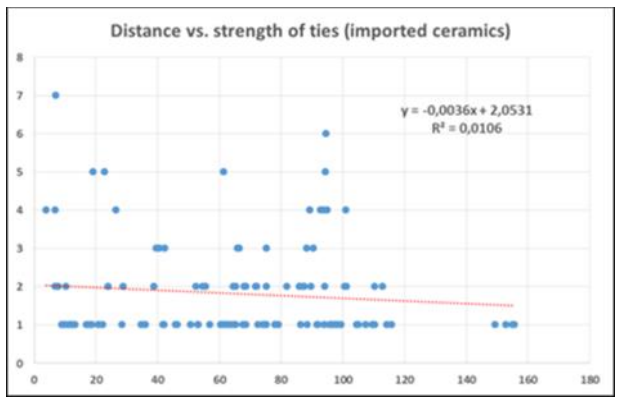

**Fig. 29** Statistical analysis of the correlation between the strength of ties between nodes and the geographical distance between them in the network model of sites on the Peloponnese connected through the co-occurrence of types of imported ceramics, 13th-15th cent. CE (data Vroom 2011; network analysis J. Preiser-Kapeller, ÖAW).

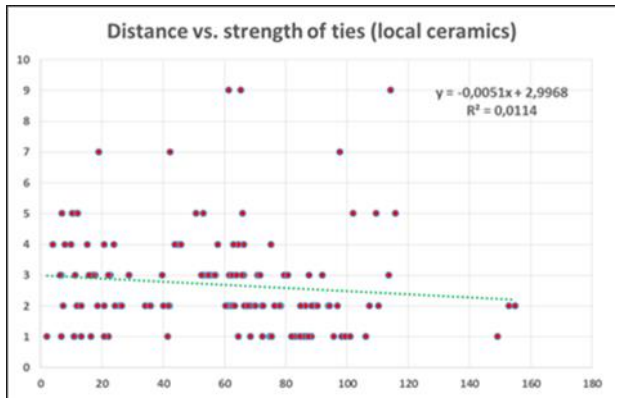

**Fig. 30** Statistical analysis of the correlation between the strength of ties between nodes and the geographical distance between them in the network model of sites on the Peloponnese connected through the co-occurrence of types of locally produced ceramics, 13th-15th cent. CE (data Vroom 2011; network analysis J. Preiser-Kapeller, ÖAW).

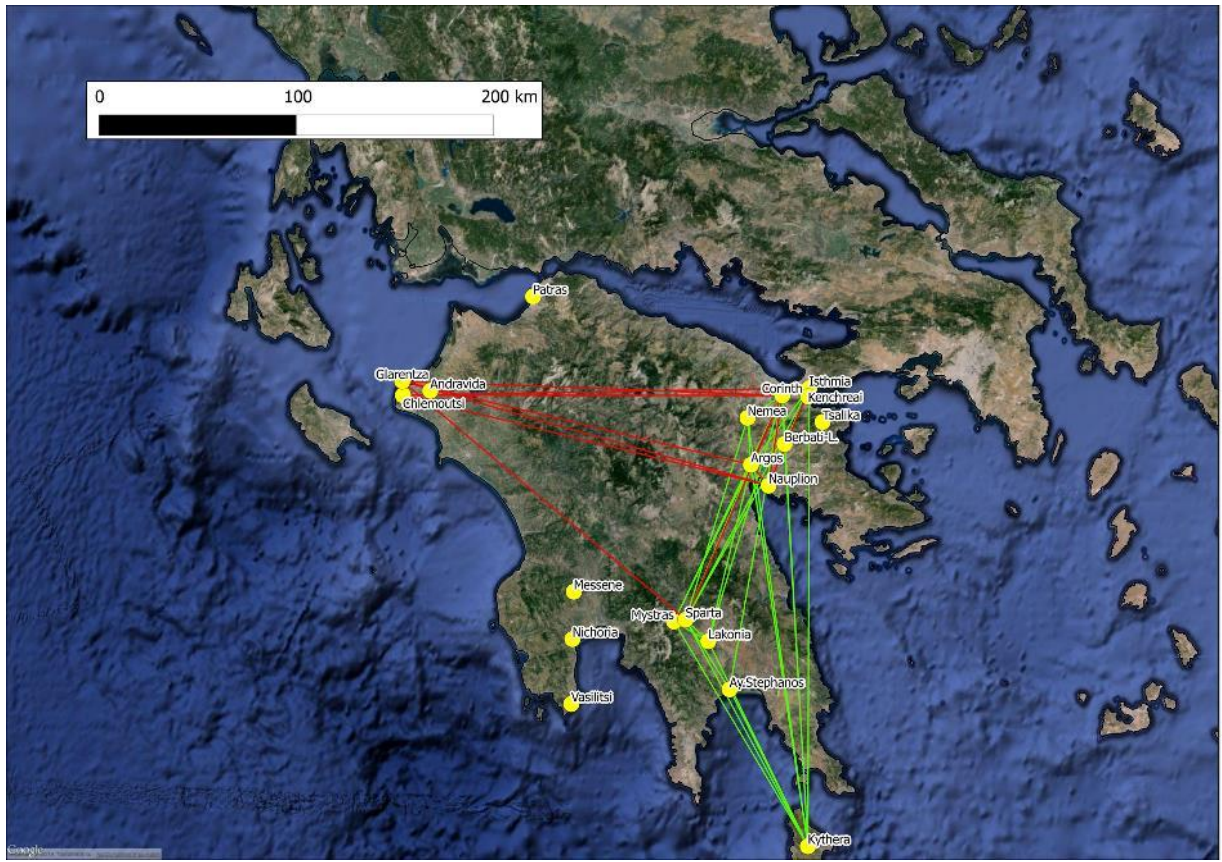

**Fig. 31** Strongest links of similarity between nodes in the network model of sites on the Peloponnese connected through the co-occurrence of types of locally produced ceramics (green) and of imported ceramics (red), 13th-15th cent. CE (data Vroom 2011; cartography and network analysis J. Preiser-Kapeller, ÖAW).

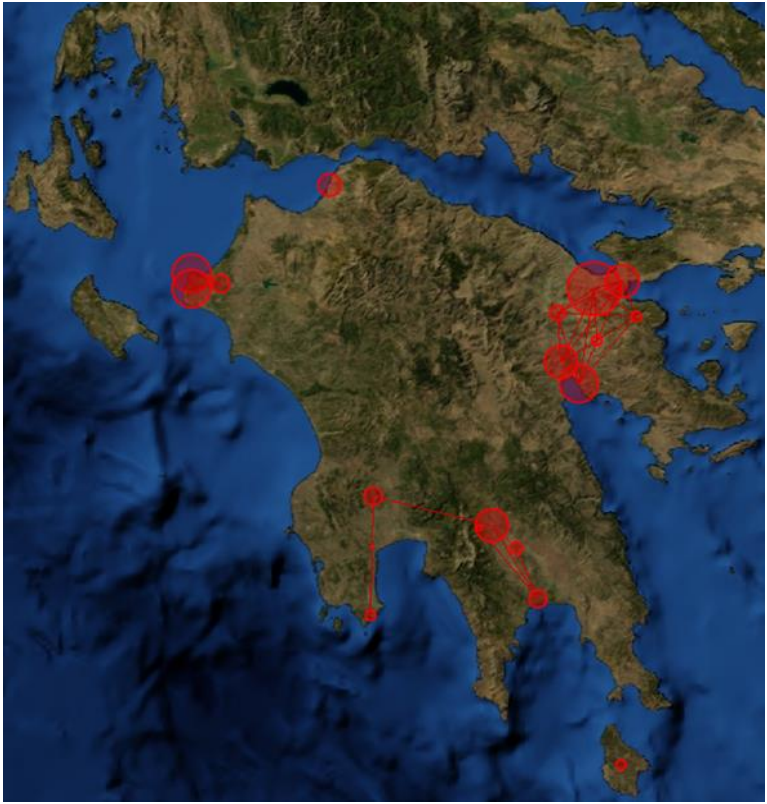

**Fig. 32** Overlay of a maximum distance network on the network model of sites on the Peloponnese connected through the co-occurrence of types of imported ceramics, 13th-15th cent. CE; cut off of all ties beyond a distance of 50 km, nodes sized according to their degree centrality values (data Vroom 2011; cartography and network analysis J. Preiser-Kapeller, ÖAW).

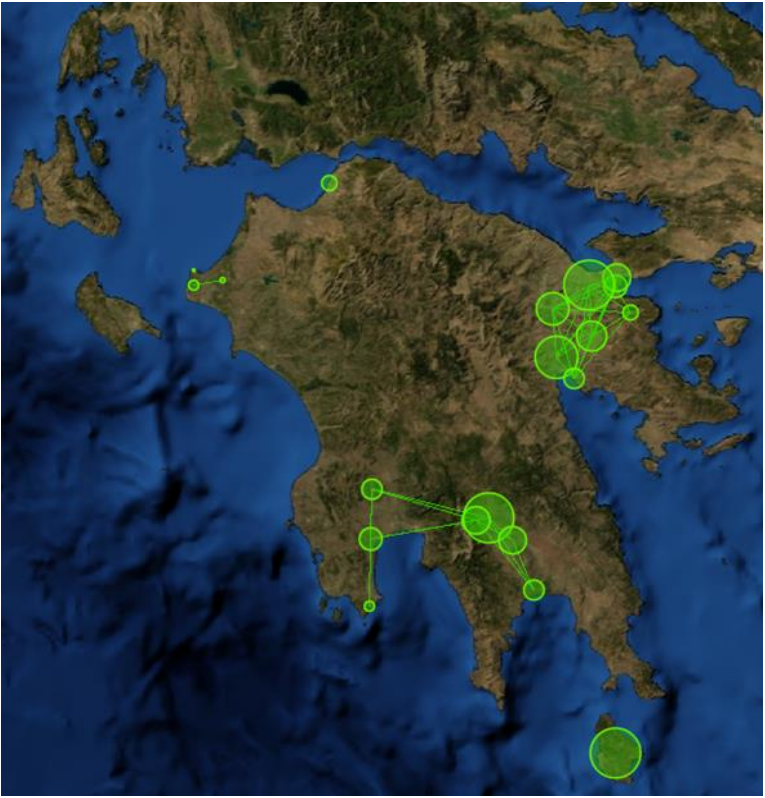

**Fig. 33** Overlay of a maximum distance network on the network model of sites on the Peloponnese connected through the co-occurrence of types of locally produced ceramics, 13th-15th cent. CE; cut off of all ties beyond a distance of 50 km, nodes sized according to their degree centrality values (data Vroom 2011; cartography and network analysis J. Preiser-Kapeller, ÖAW).

# Bibliography

Arnaud 2005: P. Arnaud, Les routes de la navigation antique. Itinéraires en Méditerranée (Paris 2005).

Barthélemy 2011: M. Barthélemy, Spatial Networks. Physics Reports 499, 2011, 1-101.

Brughmans 2012: T. Brughmans, Thinking through networks: a review of formal network methods in archaeology. Journal of Archaeological Method and Theory 20, 2012, 623-662.

Campbell 2012: J. B. Campbell, Rivers and the power of ancient Rome. Studies in the history of Greece and Rome (Chapel Hill 2012).

Carter 1969: F. W. Carter, An Analysis of the Medieval Serbian Oecumene: A Theoretical Approach, Geografiska Annaler. Series B, Human Geography, Vol. 51, No. 1, 1969, 39-56.

Collar/Coward /Brughmans/Mills: A. Collar / F. Coward / T. Brughmans / B. J. Mills, Networks in Archaeology: Phenomena, Abstraction, Representation. In: A. Collar / F. Coward / T. Brughmans / B. J. Mills (eds.), The Connected Past: critical and innovative approaches to networks in archaeology. A special issue of the Journal of Archaeological Method and Theory 22 (1) (2015) 1-31.

Cremonini/Labate/Curina 2013: S. Cremonini / D. Labate / R. Curina, The late-antiquity environmental crisis in Emilia region (Po river plain, Northern Italy): Geoarchaeological evidence and paleoclimatic considerations, Quaternary International 316, 2013, 162-178.

Delbarre-Bärtschi/Hathaway 2013: S. Delbarre-Bärtschi / N. Hathaway (eds.), EntreLacs. Das Drei-Seen-Land zur Zeit der Römer (Avenches 2013).

de Soto 2013: P. de Soto, El sistema de transportes del suroeste peninsular en época romana. Análisis de del funcionamiento de sus infraestructuras. In: J. J. Avila / M. B. Alvarez / M. G. Cabezas (eds.), Actas del VI Encuentro de Arqueología del Suroeste Peninsular. VI Encuentro de Arqueología del Suroeste Peninsular. Villafranca de los Barros, 4-6 octubre de 2012 (Huelva 2013) 1551–1576.

Ducruet/Zaidi 2012: C. Ducruet / F. Zaidi, Maritime constellations: A complex network approach to shipping and ports. Maritime Policy and Management 39, 2 (2012) 151-168.

Eckoldt 1980a: M. Eckoldt, Schiffahrt auf kleinen Flüssen Mitteleuropas in Römerzeit und Mittelalter, Schriften des Deutschen Schiffahrtsmuseums 14 (Oldenburg, Hamburg, Munich 1980).

Eckoldt 1980b: M. Eckoldt, Über das römische Projekt eines Mosel-Saône-Kanals. Deutsches Schiffahrtsarchiv 3, 1980, 29–34.

Ellmer 1984: D. Ellmers, Frühmittelalterliche Handelsschiffahrt in Mittel- und Nordeuropa, Offa-Bücher 28 (Neumünster 1984).

Escher/Hirschmann 2005: M. Escher / F. G. Hirschmann (eds.), Die urbanen Zentren des hohen und späteren Mittelalters. Vergleichende Untersuchungen zu Städten und Städtelandschaften im Westen des Reiches und in Ostfrankreich, Trierer historische Forschungen 50 (Trier 2005)

Gorenflo/Bell 1991: L. J. Gorenflo / Th. L. Bell, Network Analysis and the Study of past regional Organization. In: Ch. D. Trombold (ed.), Ancient road networks and settlement hierarchies in the New World (Cambridge 1991) 80-98.

Graßhoff/Mittenhuber 2009: G. Graßhoff / F. Mittenhuber (eds.), Untersuchungen zum Stadiasmos von Patara. Modellierung und Analyse eines antiken geographischen Streckennetzes (Bern 2009).

Gross 2012: U. Gross, Keramikgruppen des 8. bis 12. Jahrhunderts am nördlichen Oberrhein. Zur Frage von Verbreitungsgebieten und Produktionsstätten. In: L. Grunwald/H. Pantermehl/R. Schreg (Eds.), Hochmittelalterliche Keramik am Rhein. Eine Quelle für Produktion und Alltag des 9. bis 12. Jahrhunderts. RGZM-Tagungen 13 (Mainz 2012) 63–76.

Grunwald 2012: L. Grunwald, Anmerkungen zur Mayener Keramikproduktion des 9. bis 12. Jahrhunderts. In: L. Grunwald/H. Pantermehl/R. Schreg (Eds.), Hochmittelalterliche Keramik am Rhein. Eine Quelle für Produktion und Alltag des 9. bis 12. Jahrhunderts. RGZM-Tagungen 13 (Mainz 2012) 143–157.

Haase/Werther/Wunschel 2015: C. Haase / L. Werther / A. Wunschel, Güterdistribution und Verkehrsinfrastruktur klösterlicher Grundherrschaft im Frühmittelalter im Spannungsfeld ausgewählter historischer und archäologischer Quellen. In: C. Later / M. Helmbrecht / U. Jecklin-Tischhauser (eds.), Infrastruktur und Distribution zwischen Antike und Mittelalter. Tagungsbeiträge des Arbeitskreises Spätantike und Frühmittelalter 8 (Hamburg 2015) 151–189.

Heher/Preiser-Kapeller/Simeonov 2015: D. Heher – J. Preiser-Kapeller – G. Simeonov, Staatliche und maritime Strukturen an den byzantinischen Balkanküsten. In: Th. Schmidts / M. M. Vučetić (eds.), Häfen im 1. Millenium AD. Bauliche Konzepte, herrschaftliche und religiöse Einflüsse. Plenartreffen im Rahmen des DFG-Schwerpunktprogramms 1630 „Häfen von der Römischen Kaiserzeit bis zum Mittelalter" im Römisch-Germanischen Zentralmuseum Mainz, 13.-15. Januar 2014, RGZM-Tagungen 22, zugleich: Interdisziplinäre Forschungen zu den Häfen von der Römischen Kaiserzeit bis zum Mittelalter in Europa 1 (Mainz 2015) 93-116.

Isaksen 2008: L. Isaksen, The Application of Network Analysis to Ancient Transport Geography: A Case Study of Roman Baetica, Digital Medievalist, 2008: http://www.digitalmedievalist.org/journal/4/Isaksen/.

Jacoby 2013: D. Jacoby, Rural Exploitation and Market Economy in the Late Medieval Peloponnese. In: Sh. E. J. Gerstel, Viewing the Morea. Land and People in the Late Medieval Peloponnese (Washington, D. C. 2013) 213-275.

Johanek 1987: P. Johanek, Der fränkische Handel der Karolingerzeit im Spiegel der Schriftquellen. In: K. Düwel et al. (eds.), Untersuchungen zu Handel und Verkehr der vor- und frühgeschichtlichen Zeit in Mittel- und Nordeuropa, Teil IV. Der Handel der Karolinger- und Wikingerzeit. Abhandlungen der Akademie der Wissenschaften in Göttingen, Philologisch-historische Klasse 156 (Göttingen 1987) 7–68.

Kennecke 2014: H. Kennecke (ed.), Der Rhein als europäische Verkehrsachse. Die Römerzeit. Bonner Beiträge zur vor- und frühgeschichtlichen Archäologie 16 (Bonn 2014).

Knappett 2013: Ch. Knappett (ed.), Network-Analysis in Archaeology. New Approaches to Regional Interaction (Oxford 2013).

Leidwanger 2013: J. Leidwanger, Modeling distance with time in ancient Mediterranean seafaring: a GIS application for the interpretation of maritime connectivity. Journal of Archaeological Science 40, 2013, 3302-3308.

Leidwanger/Knappett et al. 2014: J. Leidwanger / C. Knappett et al., A manifesto for the study of ancient Mediterranean maritime networks, 2014, online: http://journal.antiquity.ac.uk/projgall/leidwanger342.

Lemercier 2012: Cl. Lemercier, Formale Methoden der Netzwerkanalyse in den Geschichtswissenschaften: Warum und Wie?, in: A. Müller – W. Neurath (eds.), Historische Netzwerkanalysen, Österreichische Zeitschrift für Geschichtswissenschaften 23/1 (Innsbruck, Vienna, Bozen 2012) 16–41.

Malkin 2011: I. Malkin, A Small Greek World: Networks in the Ancient Mediterranean. Greeks Overseas (Oxford, New York 2011).

McCormick 2001: M. McCormick, Origins of the European Economy. Communications and Commerce AD 300-900 (Cambridge 2001).

McCormick et al. 2013: M. McCormick et al. 2013 - Roman Road Network (version 2008), http://darmc.harvard.edu/icb/icb.do.

Mol/Hoogland/Hofman 2015: A. A. A. Mol / M. L. P. Hoogland / C. L. Hofman, Remotely Local: Ego-networks of Late Pre-colonial (AD 1000–1450) Saba, North-eastern Caribbean. In: A. Collar / F. Coward / T. Brughmans / B. J. Mills (eds.), The Connected Past: critical and innovative approaches to networks in archaeology. A special issue of the Journal of Archaeological Method and Theory 22 (1) (2015) 275-305.

Newman 2010: M. E. J. Newman, Networks. An Introduction (Oxford 2010).

Patitucci Uggeri 2005: S. Patitucci Uggeri, Il sistema fluvio-lagunare, l`insediamento e le difese del territorio revennate settentrionale (V-VIII secolo). In: Fondazione Centro Italiano di Studi sull'Alto Medioevo (ed.), Ravenna da capitale imperiale a capitale esarcale. Atti del XVII Congresso internazionale di studio sull'alto medioevo. Atti dei congressi 17,1 (Spoleto 2005) 253–360.

Pitts 1978: F. R. Pitts, The Medieval River Trade Network of Russia Revisited. Social Networks 1, 1978, 285-292.

Preiser-Kapeller 2013: J. Preiser-Kapeller, Mapping maritime networks of Byzantium. Aims and prospects of the project "Ports and landing places at the Balkan coasts of the Byzantine Empire". In: F. Karagianni (ed.), Proceedings of the conference "Olkas. From Aegean to the Black Sea. Medieval Ports in the Maritime Routes of the East" (Thessalonike 2013) 467-492.

Preiser-Kapeller 2014: J. Preiser-Kapeller, Entangling the Morea: a network model of ceramic distributions on the late medieval Peloponnese, online: http://www.academia.edu/5937949/Entangling_the_Morea_a_network_model_of_ceramic_distributions_on_the_late_medieval_Peloponnese.

Preiser-Kapeller 2015a: J. Preiser-Kapeller, Thematic introduction. In: J. Preiser-Kapeller / F. Daim (eds.), Harbours and Maritime Networks as Complex Adaptive Systems, RGZM – Tagungen 23, zugleich: Interdisziplinäre Forschungen zu den Häfen von der Römischen Kaiserzeit bis zum Mittelalter in Europa 2 (Mainz 2015) 1-24.

Preiser-Kapeller 2015b: J. Preiser-Kapeller, The Maritime Mobility of Individuals and Objects: Networks and Entanglements. In: J. Preiser-Kapeller / F. Daim (eds.), Harbours and Maritime Networks as Complex Adaptive Systems, RGZM – Tagungen 23, zugleich: Interdisziplinäre Forschungen zu den Häfen von der Römischen Kaiserzeit bis zum Mittelalter in Europa 2 (Mainz 2015) 119-140.

Preiser-Kapeller/Daim 2015: J. Preiser-Kapeller / F. Daim (eds.), Harbours and Maritime Networks as Complex Adaptive Systems, RGZM – Tagungen 23, zugleich: Interdisziplinäre Forschungen zu den Häfen von der Römischen Kaiserzeit bis zum Mittelalter in Europa 2 (Mainz 2015).

Rieth 2010: É. Rieth (ed.), Les épaves de Saint-Georges, Lyon – Ier-XVIIIe siècles. Analyse architecturale et études complémentaires. Archaenautica 16 (Paris 2010).

Rodrigue/Comtois/Slack 2013: J.-P. Rodrigue / Cl. Comtois / B. Slack, The Geography of Transport Systems (3rd ed., London, New York 2013).

Rossiaud 2007: J. Rossiaud, Le Rhône au Moyen Âge. Histoire et représentation d'un fleuve (Paris 2007).

Scheidel/Meeks et al.: W. Scheidel / E. Meeks et al., ORBIS: The Stanford Geospatial Network Model of the Roman World: http://orbis.stanford.edu/.

Schmidts 2011: T. Schmidts, Akteure und Organisation der Handelsschifffahrt in den nordwestlichen Provinzen des römischen Reiches, Monographien des Römisch-Germanischen Zentralmuseums 97 (Mainz 2011).

Sindbæk 2007: S. M. Sindbæk, The Small World of the Vikings. Networks in Early Medieval Communication and Exchange. Norwegian Archaeological Review 40, 2007, 59-74.

Sindbæk 2013: S. M. Sindbæk, Broken links and black boxes: material affiliations and contextual network synthesis in the Viking world. In: C. Knappett (ed.), Network-Analysis in Archaeology. New Approaches to Regional Interaction (Oxford 2013) 71-94.

Sindbæk 2015: S. M. Sindbæk, Northern emporia and maritime networks. Modelling past communication using archaeological network analysis. In: J. Preiser-Kapeller / F. Daim (eds.), Harbours and Maritime Networks as Complex Adaptive Systems, RGZM – Tagungen 23, zugleich: Interdisziplinäre Forschungen zu den Häfen von der Römischen Kaiserzeit bis zum Mittelalter in Europa 2 (Mainz 2015) 105–117.

Suttor 2006: M. Suttor, Vie et dynamique d'un fleuve. La Meuse de Sedan à Maastricht (des origines à 1600). Bibliothèque du Moyen Âge 241 (Bruxelles 2006).

Szabó 1984: T. Szabó, Antikes Erbe und karolingisch-ottonische Verkehrspolitik. In: L. Fenske (ed.), Institutionen, Kultur und Gesellschaft im Mittelalter. Festschrift für Josef Fleckenstein zu seinem 65. Geburtstag (Sigmaringen 1984) 125–145.

Taaffe/Gauthier, Jr. 1973: E. J. Taaffe / H. L. Gauthier, Jr., Geography of Transportation (Englewood Cliffs, N. J. 1973).

Tac. ann.: Annalen, P. Cornelius Tacitus, Ed. E. Heller (München u. a. [2]1992).

Tartaron 2013: Th. Tartaron, Maritime Networks in the Mycenaean World (Cambridge 2013).

van Lanen et al. 2015: R. J. van Lanen et al., Best travel options: Modelling Roman and early-medieval routes in the Netherlands using a multi-proxy approach. Journal of Archaeological Science: Reports 3, 2015, 144–159.

Veikou 2012: M. Veikou, Byzantine Epirus: A topography of transformation. Settlements from the 7th to the 12th centuries (Leiden, Boston 2012).

Vroom 2011: J. Vroom, The Morea and its links with Southern Italy after AD 1204: ceramics and identity. Archeologia Medievale 38, 2011, 409-430.

Wawrzinek 2009: C. Wawrzinek, Flussverlagerungen und damit einhergehende Probleme für antiken Wasserbau und archäologische Forschung. In: T. Mattern / A. Vött (eds.), Mensch und Umwelt im Spiegel der Zeit. Aspekte geoarchäologischer Forschungen im östlichen Mittelmeergebiet. Philippika 1 (Wiesbaden 2009) 171–180.

Wawrzinek 2014: Ch. Wawrzinek, In portum navigare. Römische Häfen an Flüssen und Seen (Berlin 2014).

Westerdahl 2000: C. Westerdahl, From Land to Sea, from Sea to Land. On Transport Zones, Borders and Human Space. In: J. Litwin (ed.), Down the River to the Sea. Proceedings of the Eighth International Smposium on Boat and Ship Archaeology, Gdansk 1997 (Gdansk 2000) 11–20.

Wickham 2005: C. Wickham, Framing the Early Middle Ages. Europe and the Mediterranean 400 - 800 (Oxford 2005).